\documentclass[11pt]{amsart}
\usepackage{amsmath}
\usepackage{amsxtra}
\usepackage{amscd}
\usepackage{amsthm}
\usepackage{amsfonts}
\usepackage{amssymb}
\usepackage{eucal}
\usepackage{epsfig}

\usepackage{latexsym,amsthm}

\DeclareFontEncoding{OT2}{}{}

\input epsf

\def\figin{\epsfcheck\figin}\def\figins{\epsfcheck\figins}
\def\epsfcheck{\ifx\epsfbox\UnDeFiNeD
\message{(NO epsf.tex, FIGURES WILL BE IGNORED)}
\gdef\figin##1{\vskip2in}\gdef\figins##1{\hskip.5in}
\else\message{(FIGURES WILL BE INCLUDED)}%
\gdef\figin##1{##1}\gdef\figins##1{##1}\fi}
\def\DefWarn#1{}
\def\figinsert{\goodbreak\topinsert}
\def\ifig#1#2#3#4{\DefWarn#1\xdef#1{fig.~\the\figno}
\writedef{#1\leftbracket fig.\noexpand~\the\figno}%
\figinsert\figin{\centerline{\epsfxsize=#3mm \epsfbox{#2}}}
\bigskip\medskip\centerline{\vbox{\baselineskip12pt
\advance\hsize by -1truein\noindent\footnotefont{\sl Fig.~\the\figno:}\sl\ #4}}
\bigskip\endinsert\noindent\global\advance\figno by1}

\newcommand{\dd}{{\mathrm{d}}}
\newcommand{\tr}{{\mathrm{tr}}}
\newcommand{\Det}{{\mathrm{Det}}}
\newcommand{\pa}{{\partial}}

\newcommand{\CA}{\mathcal A}
\newcommand{\CB}{\mathcal B}
\newcommand{\CC}{\mathcal C}
\newcommand{\CDd}{\mathcal D}
\newcommand{\CE}{\mathcal E}
\newcommand{\CF}{\mathcal F}
\newcommand{\CH}{\mathcal H}

\newcommand{\CJ}{\mathcal  J}

\newcommand{\CL}{\mathcal L}
\newcommand{\CM}{\mathcal M}
\newcommand{\CN}{\mathcal N}
\newcommand{\CO}{\mathcal O}
\newcommand{\CP}{\mathcal P}
\newcommand{\CQ}{\mathcal Q}

\newcommand{\CU}{\mathcal U}

\newcommand{\CW}{\mathcal W}
\newcommand{\CZ}{\mathcal Z}

\newcommand{\pb}{\bar\partial}
\newcommand{\zb}{\bar z}

\newcommand{\al}{\alpha}
\newcommand{\be}{\beta}

\newcommand{\de}{\delta}
\newcommand{\ve}{\varepsilon}
\newcommand{\vt}{\vartheta}
\newcommand{\m}{\mu}
\newcommand{\n}{\nu}
\newcommand{\la}{\lambda}
\newcommand{\s}{\sigma}
\newcommand{\om}{\omega}

\newcommand{\ba}{\bf a}

\newcommand{\bC}{\bf C}

\newcommand{\bE}{\bf E}

\newcommand{\bH}{\bf H}
\newcommand{\bI}{\bf I}
\newcommand{\bp}{\bf p}
\newcommand{\bq}{\bf q}
\newcommand{\bP}{\bf P}

\newcommand{\bR}{\bf R}
\newcommand{\bS}{\bf S}
\newcommand{\bT}{\bf T}
\newcommand{\bt}{\bf t}

\newcommand{\bZ}{\bf Z}

\newcommand{\qq}{\mathrm{q}}

\numberwithin{equation}{section}

\begin{document}

\hfill TCD-MATH-09--19

\hfill HMI-09-09

\hfill IHES/P/09/38

\title{QUANTIZATION OF INTEGRABLE SYSTEMS\\ 
\vspace{.5cm}
AND\\ 
\vspace{.5cm}
FOUR DIMENSIONAL GAUGE THEORIES}
\vspace{1cm}

\author{\sl Nikita A. Nekrasov}

\address{IHES, Le Bois-Marie,
35 route de Chartres,
Bures-sur-Yvette,  91440, France\\
Simons Center for Geometry and Physics, Stony Brook University, 
NY 11794 USA\\
E-mail: nikitastring@gmail.com}

\author{\sl Samson~L.~Shatashvili}

\address{School of Mathematics, Trinity College,
Dublin 2, Ireland\\
Hamilton Mathematics Institute, Trinity College,
Dublin 2, Ireland\\
IHES, Le Bois-Marie,
35 route de Chartres,
Bures-sur-Yvette,  91440, France\\
E-mail: samson@maths.tcd.ie}

\maketitle

\centerline{\footnotesize IHES, Le Bois-Marie, 35 route de Chartres, Bures-sur-Yvette,  91440, France}

\centerline{\footnotesize Simons Center for Geometry and Physics, Stony Brook University, NY 11794 USA}

\centerline{\footnotesize E-mail: nikitastring@gmail.com}

\bigskip

\centerline{\footnotesize  School of Mathematics, Trinity College, Dublin 2, Ireland}

\centerline{\footnotesize Hamilton Mathematics Institute, Trinity College, Dublin 2, Ireland}

\centerline{\footnotesize IHES, Le Bois-Marie, 35 route de Chartres, Bures-sur-Yvette,  91440, France}

\centerline{\footnotesize E-mail: samson@maths.tcd.ie}

\bigskip

\begin{abstract}
{We study four dimensional ${\CN}=2$ supersymmetric gauge theory in the $\Omega$-background with the two dimensional ${\CN}=2$ super-Poincare invariance. We explain how this gauge theory provides the quantization of the classical integrable system underlying the moduli space of vacua of the ordinary four dimensional
${\CN}=2$ theory. 
  The $\ve$-parameter of the $\Omega$-background is identified with the Planck constant, 
 the twisted chiral ring maps to quantum Hamiltonians, 
  the supersymmetric vacua are identified with 
  Bethe states of quantum integrable systems.
This four dimensional gauge theory in its low energy description has two dimensional twisted superpotential  which becomes the 
Yang-Yang function of the integrable system. 
We present the thermodynamic-Bethe-ansatz like formulae for 
these functions and for the spectra of commuting Hamiltonians following the direct computation in gauge theory.    
The general construction is illustrated at the examples of the many-body systems, such as
the periodic Toda chain, the elliptic Calogero-Moser system, and
their relativistic versions, for which we present a complete characterization of the
$L^{2}$-spectrum. 
We very briefly discuss the quantization of Hitchin system.}
\vskip 3cm

\end{abstract}


\bigskip

\newpage

\tableofcontents
\newpage
\section{Introduction}\label{aba:sec1}

It has been realized in the recent years \cite{MNS, GS-one, GS-two, BAsh2008-one, BAsh2008-two} that there exists an intimate connection between the vacua of the supersymmetric gauge theories and the quantum integrable systems.  This connection is quite general and applies to gauge theories in various spacetime dimensions. In the short review articles \cite{BAsh2008-one, BAsh2008-two} a large class of 
the two dimensional gauge theories were shown to correspond to the finite dimensional spin chains (and their various limits usually studied in literature on integrable models).

We report here on the new developments - we establish the connection between the four dimensional supersymmetric gauge theories and 
the quantum many-body systems. More precisely, we consider the ${\CN}=2$ supersymmetric gauge theories in four dimensions.  We subject it to the $\Omega$-background   \cite{Neksw} in two out of four dimensions. 
The $\Omega$-background is a particular background of 
 the ${\CN}=2$ supergravity in four dimensions. The general $\Omega$-background is characterized by two complex parameters 
${\ve}_{1}, {\ve}_{2}$, which have the dimension of mass. These parameters were introduced in \cite{MNS, LNStest} and used in 
\cite{MNSbound} to regularize the integrals over the instanton moduli spaces which 
arise in the supersymmetric gauge theories and the bound state problems in the
supersymmetric quantum mechanics. It was suggested in \cite{LNStest} that by deforming the Donaldson supercharge $Q$
to its equivariant version $Q + V^{\mu}_{\ve} G_{\mu}$ the instanton partition
functions would become computable (in fact, an example of an instanton integral
was proposed in \cite{LNStest}) and could ultimately test the Seiberg-Witten
solution \cite{SW} of the four dimensional ${\CN}=2$ theory. 
 This program was completed in \cite{Neksw}. It turns out that both the ideology
 and the specific examples of the integral formulae of \cite{LNStest}
 are important in our current developments related to Bethe ansatz of
 the quantum many-body systems.

When both parameters 
${\ve}_{1}, {\ve}_{2}$ of the $\Omega$-background
are non-zero  the super-Poincare invariance is broken down to a superalgebra with two fermionic and two bosonic generators corresponding to the rotations in ${\bR}^{4}$.  If one of these parameters vanishes, e.g. ${\ve}_{2} = 0$, then the resulting theory has a two dimensional ${\CN}=2$ super-Poincare invariance. This is the theory we study in the present paper. It is characterized by a single complex parameter, which we shall denote simply by $\ve$. 

Our main claim is: {\it the supersymmetric vacua of this gauge theory are the eigenstates of the quantum integrable system obtained by the quantization of the classical algebraic integrable system underlying the geometry of 
the moduli space ${\CM}_{\rm v}$ of undeformed $\CN=2$ theory. The Planck constant, the parameter of the quantization, is identified with the deformation parameter $\ve$. }

Recall that the four dimensional gauge theory with ${\ve}_1={\ve}_2=0$ 
is characterized, at low energy, by the 
prepotential ${\CF} (a)$ which 
is known to be related to some classical algebraic integrable system. For example, the pure $SU(N)$
${\CN}=2$ super-Yang-Mills corresponds to the periodic $A_{N-1}$ Toda chain \cite{GorskySW}. As shown in 
\cite{WittenDonagi} the vector multiplet part ${\CM}_{\rm v}$ of the moduli space of vacua, the Coulomb branch, of any ${\CN}=2$ supersymmetric gauge theory, is a base of the Liouville fibration of 
some classical algebraic integrable system.
Once the classical integrable system
is identified one can write down the gauge theory low energy effective action. 
When ${\ve}_1={\ve} \neq 0,\, {\ve}_2=0$ the r\^ole of $\CF(a)$ 
in the gauge theory is played, in a certain sense, by the two dimensional
twisted superpotential ${\CW}(a ; {\ve})$.  We show it also 
has a meaning in the algebraic integrable system,  
albeit in the quantum one.  It is identified with the
Yang-Yang counting function \cite{YangYang} governing the spectrum of the quantum system. 
One does not always know a priori how to quantize an algebraic integrable 
system, just knowing its classical version does not suffice. 
However,  ${\CW}(a ; {\ve})$ is computable by the gauge theory methods
(and so is ${\CF}(a)$), so the situation got reversed - 
the gauge theory helps to learn about the quantization of an integrable system.

This is an important part of the general program,  whose details will appear in the longer version \cite{Tobe} which in addition (and in particular) will combine the results reported here and those  in  \cite{BAsh2008-one, BAsh2008-two} to form a unified picture.

There are numerous applications of 
the correspondence 
 \cite{MNS, GS-one, GS-two, BAsh2008-one, BAsh2008-two}: 
the gauge theory applications, the study of quantum 
cohomology, (infinite-dimensional) representation theory, harmonic analysis,
the many-body
quantum mechanics.

In a sense the most general non-relativistic algebraic
integrable system is the so-called Hitchin system. 
Its quantization is an interesting and important 
problem, whose special cases, corresponding to the
degenerate Riemann surfaces, are by now old and classical
problems. In this paper we  illustrate the power of
our methods at the examples of two such degenerations,
the quantum elliptic Calogero-Moser system, 
and its limit, the periodic Toda chain. 
We shall present a complete characterization of the
$L^{2}$-spectrum of these systems.

The paper is organized as follows. The section $\bf 2$ reviews the correspondence
between the supersymmetric vacua of gauge theories with two dimensional
${\CN}=2$ supersymmetry and Bethe vectors of some quantum integrable systems,
and lays down a route to the analysis of the four dimensional theories. The
section $\bf 3$ is devoted to the four dimensional theories, the $\Omega$-background, the $F$-terms, and the calculation of the effective twisted superpotential. 
The section $\bf 4$ reviews the integrability side of the story, first the classical algebraic
integrable systems, then their quantization. The section $\bf 5$ describes the
construction at the examples of the periodic Toda chain, elliptic Calogero-Moser
system, the Ruijsenaars-Schneider model, and the Hitchin system. The section $\bf 6$ presents the thermodynamic Bethe ansatz-like formulae for ${\CW}(a, {\ve})$. 

{\bf Acknowledgments:} The research of NN was supported by  {\it l'Agence Nationale de la Recherche} under the grants
ANR-06-BLAN-3$\_$137168 and ANR-05-BLAN-0029-01, by the Russian Foundation for Basic Research through the grants
RFFI 06-02-17382 and NSh-8065.2006.2, that of SSh was supported by SFI grants 05/RFP/MAT0036 and 08/RFP/MTH1546,  by European Science Foundation grant ITGP and by the funds of the Hamilton Mathematics Institute TCD.

We thank S.~Frolov, A.~Gerasimov, A.~Gorsky, 
A.~Okounkov, E.~Sklyanin, F.~Smirnov,  J. Stalker, 
L.~Takhtajan, P.~Wiegmann and E.~Witten for the discussions.

This paper is the expanded version of the plenary 
talk given by one of us (SSh)
at the International Congress of Mathematical Physics, Prague, August 2009. We have also reported
the results of this paper at Strings'09 (Rome), 
Okunfest (Moscow), at the 
Symposium on Theoretical and Mathematical Physics 
(St. Petersburg), at the conference on 
ÒIntegrability in Gauge and String TheoryÓ (Potsdam), ENS summer
institute (Paris). 
We are grateful
to the 
organizers for the invitations and for the opportunity to present our
results there, and to 
the audiences and the participants for their questions.

\newpage
\section{Review of the Bethe/gauge correspondence}

In this section we remind the correspondence between
the supersymmetric vacua of the two dimensional
${\CN}=2$ theories and the
stationary (eigen)states of quantum integrable
systems and then propose a four dimensional 
generalization. The correspondence can also be
applied to the study of the topological
field theory  obtained by twisting the physical supersymmetric
gauge theory. 

\subsection{Twisted chiral ring and quantum integrability}

The space of supersymmetric vacua of a theory with four supersymmetries carries a representation of a commutative associative algebra, the so-called (twisted) chiral ring,  see e.g book \cite{HoriBook}. 

For example, in two space-time dimensions, the ${\CN}=2$ supersymmetry is generated by the fermionic charges $Q_{\pm}, {\bar Q}_{\pm}$, which obey the anticommutation relations:
\[
\{ Q_{\pm} , {\bar Q}_{\pm} \}  =  2 ( H \pm P ) \, ,
\]
\begin{equation}
\{ Q_{+}, Q_{-} \} =  \{ {\bar Q}_{+}, {\bar Q}_{-} \}  =  0  \,
\label{eq:susy} 
\end{equation}
\[
\{ Q_{+}, {\bar Q}_{-} \} =  \{ {\bar Q}_{+},  Q_{-} \}  =  0  \, \]
the last two lines being valid in the absence of the central extension, induced, e.g. by some global symmetry charges.

The twisted chiral ring is generated by the operators ${\CO}_{k}$, $k = 1, 2, \ldots$ which (anti)-commute with the 
operator
\begin{equation}
{\CQ}_{A} = Q_{+} + {\bar Q}_{-}\, , \qquad \{ {\mathcal Q}_A, {\mathcal Q}_A^{\dagger} \}=H
\label{eq:harmo}
\end{equation}
Analogously one defines the chiral ring, whose generators (anti)commute with the operator:
\begin{equation}
{\CQ}_{B} = Q_{+} + Q_{-}\, , \qquad \{ {\CQ}_B, {\CQ}_B^{\dagger} \}=H
\label{eq:harmoo}
\end{equation}
In our work we concentrate on the ${\CQ}_{A}$-cohomology and assume that the possible central extension of (\ref{eq:susy}) leaves ${\CQ}_{A}$ nilpotent ${\CQ}_{A}^{2} = 0$. The local operators ${\CO}_{k}(x)$ are independent up to the 
${\CQ}_{A}$-commutators of their location $x$. Their operator product expansion defines a commutative associative ring, 
\begin{equation}
{\CO}_{i} {\CO}_{j} = c_{ij}^k {\CO}_{k} + 
\{ {\CQ}_{A}, \ldots \}
\label{chiral}
\end{equation}
If $|0 \rangle$ is a vacuum state of the Hamiltonian $H$, $H | 0 \rangle = 0$, then so is ${\CO}_i |0 \rangle = |i \rangle$, and moreover the space of vacua is the representation of the
twisted chiral ring.

Thus the space of supersymmetric vacua, which can be effectively studied using the cohomology of the operator ${\CQ}_{A}$ (or ${\CQ}_B$), is the space of states of some quantum integrable system:
\begin{equation}
{\bH}^{\rm quantum} = {\rm ker} {\CQ}_{A} / {\rm im} {\CQ}_{A}
\label{eq:susyvac}
\end{equation}
The actual vacua are the harmonic representatives of this cohomology as follows from (\ref{eq:harmo}), (\ref{eq:harmoo}). The operators ${\CO}_{k}$ and more generally the functions of ${\CO}_{k}$'s are the quantum Hamiltonians.

The operators ${\CO}_{k}$ can be chosen {\sl under the assumption of the absence of massless charged matter fields }
to be the gauge invariant polynomials of the complex scalar $\sigma$ in the vector supermultiplet, 
\begin{equation}
{\mathcal O}_{k} = {1\over k! (2\pi i )^k} 
{\mathrm{Tr}}\, {\sigma}^{k}  
\end{equation}
One is looking for the common eigenstates of these Hamiltonians:
\begin{equation}
{\bf\Psi}_{\la} \in {\bH}^{\rm quantum}\, , \qquad {\CO}_{k}
{\bf\Psi}_{\la} = E_{k}({\la}) {\bf\Psi}_{\la}
\label{eq:commeig}
\end{equation}
where $E_{k}({\la})$ are the corresponding eigenvalues, and $\la$ are some labels. In general they are complex, $E_{k}({\la}) \in {\bC}$. 

{\it The important, or at least the interesting, problem is to identify the quantum integrable system given an ${\CN}=2$ gauge theory, or to solve the converse problem -- to find the ${\CN}=2$ theory given a quantum integrable system.} 
For a large class of models on both sides this problem has been solved in \cite{MNS, GS-one, GS-two, BAsh2008-one, BAsh2008-two}.

In most interesting cases the supersymmetric gauge theory at low energies has an effective two dimensional abelian gauge theory description  (with four supercharges), the so-called theory on the Coulomb branch. The supersymmetric vacua are determined in terms of the exactly calculable effective twisted superpotential 
${\tilde W}^{\rm eff}(\sigma)$. Loosely speaking, given a vector of electric fluxes $( n_{1}, \ldots , n_{r})$, with $n_{i} \in {\bZ}$ the vacua are given by the critical points of the shifted superpotential ${\tilde W}^{\rm eff}(\sigma) - 2\pi i \sum_{i=1}^{r} n_{i} {\s}^{i}$ (sometimes we loosely refer to the 
critical points of the superpotential without explicitly saying ``shifted"):
\begin{equation}
{1\over 2\pi i}\frac{{\pa}{\tilde W}^{\rm eff}(\sigma)} {{\pa}{\s}^i} =  n_i 
\label{eq:main}
\end{equation}
as follows from the consideration of the effective potential
\begin{equation}
U_{\vec n}(\sigma) = {1 \over 2}
{\rm g}^{ij} \left( -2\pi i n_i + {{\partial \tilde{W}}^{\rm eff} \over {\partial {\sigma}^i}} \right) \left( + 2\pi i n_j + {{\partial \tilde{\bar W}}^{\rm eff} \over {\partial {\bar\sigma}^j}} \right)
\label{eq:poten}
\end{equation}
Equivalently:
\begin{equation}
{\exp} \left( {\partial {\tilde W}^{\rm eff}}({\sigma)} \over {\partial \sigma}^{i}  \right) =  1
\label{eq:bethegood}
\end{equation}

\subsection{Topological field theory}

The two dimensional
${\CN}=2$ gauge theory can be 
topologically twisted. In the twisted version the supercharge 
${\CQ}_A$ plays the r\^ole of the BRST operator. The vacua of
the physical theory are the physical states of the topological
theory. 

The action of the topological field theory
can be brought to the simple form
\begin{equation}
S= \ \sum_{i=1}^{r} \int_{\Sigma}\,
{{\pa {\tilde W}^{\rm eff}(\sigma)} \over 
{\pa \sigma_i}} F_{A}^{i}+ \frac{1}{2}
\sum_{i,j=1}^{r} \int_{\Sigma}\,
{{\pa^2 {\tilde W}^{\rm eff}(\sigma)} \over 
{\pa \sigma_i}{\pa \sigma_j}}
\psi^{i}\, \wedge \psi^{j} 
\label{eq:top}
\end{equation}
by the so-called quartet mechanism (one adds the anti-twisted 
superpotential ${\bar t}\, {\mathrm Tr} {\bar\s}^2$, and sends
${\bar t} \to \infty$).  
Here $F_A^{i}$ is a curvature of abelian gauge field $A^i$ and 
$\psi^i$ is anti-commuting $1$-form on $\Sigma$, the 
super-partner of $A^{i}$. In this form the gauge theory becomes
a contour integral over the ${\s}$ field (the field ${\bar\s}$ being eliminated). 
The original supersymmetric theory
contains information hidden in the so-called $D$-terms, which 
ultimately leads to the wall-crossing
phenomena \cite{LNS}.

The canonical quantization of the theory (\ref{eq:top}) 
on the cylinder ${\Sigma} = {\bR} 
\times {\bS}^1$ is simple. Indeed, the only physical degree of freedom
of this theory is the monodromy 
${\exp} \, i {\vartheta}^{i} = {\exp} \oint_{{\bS}^1} A^{i}$
 of the gauge field
around the circle ${\bS}^1$ and the momentum
 conjugate to $\vartheta^{i}$, 
$I_{i} = {{\pa} {\tilde W}^{\rm eff}}/{{\pa} {\s}_i}$. 
Since ${\vartheta}^i$ takes values in a circle (due to the large gauge 
transformations) the conjugate variable $I_{i}$
is quantized, leading to the equations (\ref{eq:main}), 
(\ref{eq:bethegood}).

{\it Our conclusion is that one can study the vacuum sector of the 
$\CN=2$ gauge theory or one can study
the topologically twisted version of the same theory  -- 
ultimately one deals with the Bethe states of 
some quantum integrable system. This correspondence benefits 
all three subjects involved.}

\subsection{Yang-Yang function and quantum 
spectrum}

In \cite{GS-one, GS-two} the form (\ref{eq:top}) of the effective action and the quantization argument above for the
two dimensional theory studied in \cite{MNS} 
was used to identify ${\tilde W}^{\rm eff}({\s})$ with the Yang-Yang function of the $N$-particle
Yang integrable system. However both the quantization 
argument and the identification of 
${\tilde W}^{\rm eff}$ with a Yang-Yang function are not restricted to this case
only. 

It is a remarkable feature of the quantum integrable systems that
their spectrum can be sometimes studied using
Bethe ansatz. The spectrum of the quantum system
is determined by the equations on the (quasi)momentum variables (rapidities)
${\la}_{i}$, which enter the parametrization of 
the eigenfunctions. It is even more remarkable
that the quasimomenta are determined by the
equations which have a potential:
\begin{equation}
\frac{1}{2\pi i} 
\frac{\pa Y({\la})}{\pa {\la}_{i}} =  n_{i} \in {\bZ}
\label{eq:baeq}
\end{equation}
The function $Y({\la})$ is called the counting, 
or YY function \cite{YangYang}, and (\ref{eq:baeq}) is the corresponding Bethe equation \cite{Bethe}.

It has been demonstrated in \cite{BAsh2008-one, BAsh2008-two} that for many interesting gauge theories in two, three (compactified on ${\bS}^{1}$) or four (compactified on ${\bT}^2$)  dimensions the equation (\ref{eq:bethegood}) coincides with Bethe equation, for  a large class of interesting quantum integrable systems.
Moreover, the effective twisted superpotential of 
the gauge theory equals 
the  YY function of the quantum integrable system (e. g. for $G=U(N)$):
\begin{equation}
{\tilde W}^{\rm eff} ({\s})= Y ({\la})
\label{eq:rela}
\end{equation} 
when the Coulomb branch moduli ${\s}_{i}$ are identified
with the spectral parameters ${\lambda}_{i}$ of the quantum integrable system.  
 In addition, the expectation values of the twisted chiral ring operators 
${\CO}_{k}$ in the vacuum $| {\lambda} \rangle$
given by the solution $\sigma = {\la}$ of (\ref{eq:main}) coincide with the eigenvalues of quantum Hamiltonians of integrable system:
\begin{equation}
\langle {\la} | {\CO}_{k} | {\la} \rangle = {\CE}_k ({\la})
\label{eq:os}
\end{equation}
\begin{equation}
{\bH}_{k} {\bf\Psi}_{\la} = {\CE}_k ({\la}) {\bf\Psi}_{\la}
\label{eq:ham}
\end{equation}
Most of these models of \cite{BAsh2008-one, BAsh2008-two} are intrinsically quantum. Their Hilbert spaces are finite dimensional. The Planck constant in these cases is not a continuous variable. These theories are not very natural from the point of view of the quantization procedure, where one is given a classical integrable system and is asked to construct the quantum integrable system.

\subsection{Quantization from four dimensions}

In the current paper we study a novel type of theories, which originate in four dimensions. They have a continuous parameter $\ve$, which becomes the continuous Planck constant of a
quantum integrable system. The corresponding quantum integrable systems have infinite-dimensional Hilbert spaces. For example, we shall give a solution to the quantum periodic Toda chain (pToda), an elliptic Calogero-Moser system (eCM) and their relativistic analogues.  We shall also make some remarks on the general Hitchin system. 

Our strategy is the following:
\begin{enumerate}

\item Start with a four dimensional ${\CN}=2$ gauge theory (for example, one may take a pure ${\CN}=2$ theory with some gauge group $G$ or the ${\CN}=2^{*}$ theory, the theory with one massive hypermultiplet in the adjoint representation); its low energy effective Lagrangian is determined in terms of a single multi-valued analytic function of the Coulomb moduli $(a_{1}, \ldots, a_{r})$, $r = {\rm rank}(G)$, called the prepotential, 
${\CF}(a; m, {\tau})$. Here $m, {\tau}$ etc. are the parameteres
 of the gauge theory, whose meaning will be clarified later on.

\item The theory in the ultra-violet has the observables 
${\CO}_{k}$ \[ 
{\CO}_{k} = \frac{1}{(2\pi i)^k k!} {\mathrm Tr}
{\phi}^k
\]
where $\phi$ is the complex scalar in the vector multiplet. The observables ${\CO}_{k}$
correspond to the holomorphic functions $u_{k}$ on 
${\CM}_{\rm v}$. These observables are singled out by their
(anti-)commutation with the supercharge ${\CQ}_A$ (this 
supercharge becomes the $Q$-operator of the Donaldson-Witten
theory in the standard ${\CN}=2$ twist). One can
formally deform the theory in the ultra-violet, 
\begin{equation}
{\CF}^{\rm tree}
\to {\CF}^{\rm tree} + \sum_{k} t_{k} {\CO}_{k}
\label{eq:tdef}
\end{equation} 
In 
the low energy the deformed theory is given by the family
 of prepotentials ${\CF}(a; {\bt}; m, {\tau})$, ${\bt} =
 ( t_{k} )$. 
 For some models the deformed prepotentials were computed in \cite{LNS, LMN, MN}. For our purposes the $t_{k}$ deformation can be studied formally, i.e. without addressing the convergence issues. However, some of the couplings in $\bt$ can be interpreted as shifting the ultraviolet complexified gauge coupling $\tau$. For these couplings a finite deformation can be studied. For other couplings only the first order deformation is needed (and therefore the contact term problem \cite{LNS} does not arise). The deformed prepotential 
 ${\CF}(a; {\bt}; m, {\tau})$ generates the vevs of ${\CO}_{k}$'s:
 \[ 
 \frac{{\pa}{\CF} ( a; {\bt}; m , {\tau})}{{\pa}t_{k}} = 
 \langle  {\CO}_{k} \rangle_{a}
 \]
 
\item The prepotential ${\CF}(a, m, {\tau})$ 
has an interpretation in terms of a classical algebraic integrable system where it plays the r\^ole of the generating function of a 
Lagrangian submanifold $L$ of a symplectic complex  vector space relating two types of complex action variables, 
$a^{i}$ and $a_{D,i}$. The ${\bt}$-couplings deform $L$
to another Lagrangian submanifold $L_{\bt}$ \cite{LNS}.

\item Finally, we introduce one more deformation parameter, $\ve$, which corresponds to subjecting the theory to the $\Omega$-background ${\bS}^{1} \times {\bR}^{1} \times {\bR}^{2}_{\ve}$, or  ${\bR}^{2} \times {\bR}^{2}_{\ve}$ (see the next subsection). The $F$-terms of the low energy Lagrangian are now two dimensional with the twisted superpotential which we denote by ${\CW}(a; {\bt}; \ve)$. 
For the discussion of the supersymmetric vacua only these twisted $F$-terms are relevant, therefore the effective  description of the
low energy physics is two dimensional. In particular, one derives the equation determining the vacua, as
\begin{equation}
{{\pa} {\CW}(a;{\bf t}; {\ve}) \over
{\pa}a_{i}} = 2\pi i n_{i}\, , \qquad n_{i} \in {\bZ}
\label{eq:betbao}
\end{equation}
while the spectrum of the twisted chiral observables is computed from:
\begin{equation}
{\CE}_{k} = 
{{\pa}{\CW}(a;{\bt};{\ve}) \over {\pa}t_{k}} 
\Biggr\vert_{{\bt}=0}
\label{eq:cecho}
\end{equation}
We shall note that the equation (\ref{eq:betbao}) can be written, in the examples studied in this paper and probably in more general situations, in terms of
the factorized $S$-matrix of the associated ``hyperbolic" many body system corrected by the ``finite-size" terms (i.e. the instanton corrections in ${\qq}$ in the case of ${\CN}=2^{*}$ theory, the corrections in $\Lambda$ for the pure ${\CN}=2$ theory etc.):
\begin{equation}
{{\pa} {\CW}(a;  {\ve}) \over
{\pa}a_{k}} = i {{\tau a_k} \over {\ve}}+\sum_{j \neq k} {\rm log} \, S (a_{k}-a_{j}) + \, {\rm the \ finite \ size\ terms}
\label{eq:finite}
\end{equation}
where we set $t_{k}=0$ for $k > 2$. 
\item As in 
\cite{MNS, GS-one, GS-two, BAsh2008-one, BAsh2008-two} the equations (\ref{eq:betbao}), (\ref{eq:cecho})
determine the spectrum of the quantum integrable system, where  
$\ve$ plays the r\^ole of the (complexified) Planck constant. We identify this system with the quantization of the classical algebraic
integrable system describing the low energy effective theory
of the four dimensional ${\CN}=2$ theory.

\end{enumerate}

{\bf Remark.} Note the unfortunate notational conflict. The order parameters, ${\s}_{i}$, the eigenvalues of the complex scalar in the vector multiplet,
are denoted traditionally by ${\s}_{i}$ in the context of two dimensional
gauge theories, by $a^{i}$ in the context of four dimensional
gauge theories, by ${\phi}_{i}$ in the context of topological
gauge theories. The same parameters exhibit themselves as the Bethe
roots in our correspondence with the quantum integrable systems. 
In that world they are denoted by ${\la}_{i}$. In
the context of the periodic Toda chain these variables are
denoted by ${\de}_{i}$ in \cite{KL}, and by $t_{\al}$
in \cite{FV}. We failed
to propose a unified notation and followed the traditional way
of denoting the same object by different letters corresponding to
the different context. We hope the reader will not be too confused
by this. 

\newpage
\section{Four dimensional gauge theory}

In this section we describe in some detail the $\Omega$-deformation of the supersymmetric gauge theory. We analyze the effective theory and observe that it can be related, at the level of
cohomology of some supercharge, to a supersymmetric gauge theory in two dimensions. The two dimensional theory has an effective twisted superpotential which we analyze. 

\subsection{The $\Omega$-background and twisted masses}

A quantum field theory in $k+2$ space-time dimensions can be viewed, formally, as a two dimensional theory with an infinite number of fields. If this theory is studied on flat $k+2$ dimensional space-time, or on a space-time fibered over a two-dimensional manifold ${\Sigma}$ with the Euclidean fibers ${\bR}^{k}$, it has (in addition to the global symmetries of the $k+2$-dimensional theory) a global symmetry group ${\bE}(k)$ of isometries of the $k$-dimensional Euclidean space 
${\bR}^{k}$. Accordingly, if the theory has an ${\CN}=2$ supersymmetry on $\Sigma$, then 
one may deform it by turning on the twisted masses corresponding to the global symmetry ${\bE}(k)$. Unlike the conventional global symmetries which typically form  a compact Lie group with the unique, up to a conjugation, maximal torus, the group 
${\bE}(k)$ has several, for $k > 1$, inequivalent Cartan subgroups. 
Thus there exists several physically inequivalent deformations of the $k+2$ dimensional theory. 

For example, one may choose a subgroup ${\bR}^{k-2l} \times SO(2)^{l}$ of translations in ${\bR}^{k-2l}$ and rotations in the $l$ orthogonal two-planes. The theory with twisted masses
corresponding to the translations in
${\bR}^{k-2l}$ is equivalent to the ordinary Kaluza-Klein
compactification on a torus ${\bT}^{k-2l}$. Such theories are
studied in \cite{BAsh2008-one, BAsh2008-two}. It is shown there, that if one starts with the four dimensional gauge theory with ${\CN}=2$ supersymmetry, with $L$ hypermultiplets in the fundamental representation, and the gauge group $U(N)$, then, upon
the compactification on the two-torus ${\bT}^{2}$ one gets, via
(\ref{eq:susyvac}) the $su({2})$ XYZ spin chain, where the elliptic
parameter of the spin chain is identified with the complex structure modulus $\tau$
of the compactification torus. The three dimensional theory gives rise to the XXZ spin chain, and the two dimensional one corresponds to the XXX spin chain. The number $N$ of colors
is equal to the excitation level of the spin chain (the total spin's projection $S_{z}$ is equal to $N - \frac{L}{2}$), while the number of flavors $L$ is the length of the spin chain. 

\subsubsection{Four dimensional theory on ${\bR}^2_{\ve}$}

Another possibility is to consider the twisted masses corresponding to the rotational symmetry. In this case one gets the theory in the  $\Omega$-background \cite{Neksw}.  

Consider a four dimensional
${\CN}=2$ theory on a four manifold $M^{4}$ fibered over a two dimensional base $\Sigma$ with the $\Omega$-background 
along the fibers ${\bR}^{2}$. Somewhat schematically we shall denote the fibers by ${\bR}^{2}_{\ve}$. The base $\Sigma$ of the fibration could be a two-plane ${\bR}^{1,1}$ or a cylinder ${\bS}^{1} \times {\bR}^{1}$. One can also study the twisted
theory for which $\Sigma$ could be an arbitrary Riemann surface. 

In our main examples, corresponding to the periodic Toda and elliptic Calogero-Moser systems our staring point would be the pure ${\CN}=2$ super-Yang-Mills theory, or the 
${\CN}=2^{*}$ theory, corresponding to the gauge theory with a single adjoint massive hypermultiplet. In the limit of vanishing
mass of the adjoint matter fields 
the latter becomes an ${\CN}=4$ theory. 

Now let us give some details on the $\Omega$-deformation. Let us denote the coordinates on the fiber ${\bR}^{2}_{\ve}$ 
by $(x^{2}, x^{3})$, and the coordinates on 
the base $\Sigma$ by $(x^{0}, x^{1})$. Introduce the vector field  
\begin{equation}
\label{eq:rot}
U=  x^{2}{\pa}_{3}-x^{3}{\pa}_{2}
\end{equation}
generating the $U(1)$ rotation in ${\bR}^{2}$. Let ${\ve} \in {\bC}$ be a complex parameter and let $V = {\ve} U, {\bar V} = {\bar\ve} U$ be the complex vector fields on ${\bR}^{2}$. 

The  bosonic part of the pure ${\CN}=2$ super-Yang-Mills Lagrangian of the theory on ${\bR}^{2}_{\ve}$ is simply:
\begin{equation}
L=   -{1 \over 4g_{0}^{2}} {\tr}\, F \wedge \star F + {\tr}\, \left( D_A\phi - {\ve} \iota_{U} F \right) \wedge \star \left(
D_A{\bar \phi} - {\bar\ve} \iota_{U}F \right)+
\label{eq:bosomglag}
\end{equation}
\[
+ \frac{1}{2} {\tr} \left( [\phi,{\bar \phi}]+ {\iota}_U  D_A \left( {\ve} {\bar \phi} - {\bar\ve} {\phi} \right) \right)^2 +{\theta_0 \over {2\pi}} {\tr} F \wedge F \]
It is clear that the Poincare invariance in the $(x^{0}, x^{1})$ directions  is unbroken, and it is possible to show that in fact the two dimensional ${\CN}=2$ super-Poincare invariance is preserved. Thus there are four supercharges. 

The only (twisted) $F$-terms of the low energy effective theory are two-dimensional and can be represented as the 
 non-trivial twisted superpotential ${\CW}(a; \ve)$ 
 (where as before $a$ denotes the  complex  scalar in abelian vector multiplet). 

If we send ${\ve}$ back to zero then the 
 low energy theory is fully four-dimensional with 
 a continuous moduli space ${\CM}_{\rm v}$
 of vacua and the low energy effective Lagrangian 
 is described in terms of the prepotential  ${\CF} (a)$. 
 
For non-zero $\ve$ the theory has a discrete set of vacua 
given by the minima of the potential  (\ref{eq:poten}), (\ref{eq:bethegood}).
These vacua are therefore the solutions to the equation (\ref{eq:betbao}),
which we shall later on identify with the Bethe equation of some
quantum integrable system. 

\subsubsection{Calculation of the twisted superpotential}

Our discussion would have had a rather limited 
significance were it not for the possibility of exact 
computation of ${\CW}(a ; {\ve})$.

Let us briefly explain our strategy. 
Consider the four dimensional theory in the general
$\Omega$-background, with both rotation parameters
${\ve}_1$, ${\ve}_2$ non-zero. Then the effective theory
has a prepotential ${\CF}(a, {\ve}_{1}, {\ve}_{2})$ which is analytic
in ${\ve}_{1}, {\ve}_{2}$ near zero and becomes exactly
the prepotential of the low energy effective four dimenional theory
in the limit ${\ve}_{1}, {\ve}_{2} \to 0$:
\begin{equation}
S^{\rm eff}_{4d} = \int {\CF}^{(4)} (a, {\ve}_{1}, {\ve}_{2}) + \{ Q, \ldots \}
\label{eq:effacfd}
\end{equation}
where we denote collectively by $a$ all the vector multiplet scalars
as well as background scalars, such as the masses of matter fields. 
At the same time, had we started with a two dimensional theory
which is characterized by some twisted superpotential ${\CW}(a)$, 
and had we subjected it to the two dimensional ${\Omega}$-background
with the parameter ${\ve}_{2}$, the effective action would have had the form:
 \begin{equation}
S^{\rm eff}_{2d} = \int {\CW}^{(2)} (a, {\ve}_{2}) + \{ Q, \ldots \}
\label{eq:effactd}
\end{equation}
where ${\CW} (a, {\ve}_{2}) \to {\CW}(a)$ as ${\ve}_{2} \to 0$,
and $Q$ is a certain supercharge which we use to study the vacuum states
of our theory. In the Eqs. (\ref{eq:effacfd}), (\ref{eq:effactd}) the notations $F^{(4)}$, $W^{(2)}$ refer to the cohomological descendents
of the local operators $F$, $W$, etc. 
Now the standard manipulations with equivariant cohomology
give:
\begin{equation}
\int_{{\bR}^{4}} {\CF}^{(4)} (a, {\ve}_{1}, {\ve}_{2})  = 
{1\over {\ve}_{1}} \int_{{\bR}^{2}} {\CF}^{(2)} (a, {\ve}_{1}, {\ve}_{2}) 
= {1\over {\ve}_{1}{\ve}_{2}} {\CF}(a, {\ve}_{1}, {\ve}_{2})
\label{eq:prepsup}
\end{equation}
modulo $Q$-exact terms at each step. By carefully manipulating the $Q$-exact terms
we can connect the computation of the gauge theory partition function
in the ultraviolet, which is given by a one-loop perturbative and a series
of exact instanton corrections,  to the computation in the infrared,
using the Wilsonian effective action, where the effective energy scale
can be sent all the way to zero:
\begin{equation}
{\exp}\, {1\over {\ve}_{1}{\ve}_{2}} {\CF}(a, {\ve}_{1}, {\ve}_{2}; 
{\qq})
= {\CZ} (a, {\ve}_{1}, {\ve}_{2}; {\qq}) = 
{\CZ}^{\rm pert} (a, {\ve}_{1}, {\ve}_{2}; {\qq})
\times {\CZ}^{\rm inst} (a, {\ve}_{1}, {\ve}_{2}; {\qq})
\label{eq:partfn}
\end{equation}
where we restored the dependence on the complexified bare gauge coupling
\[ 
{\qq} = {\exp}\, 2\pi i \tau\, \qquad {\tau} = {{\vartheta}\over 2\pi}
+ {4\pi i\over {\rm g}^2}
\]
Now, by comparing the Eqs. (\ref{eq:effacfd}), (\ref{eq:effactd}), (\ref{eq:partfn})
we conclude:
\begin{equation}
{\CW}(a ; {\ve} ; {\qq} ) = 
{\rm Limit}_{{\ve}_{2} \to 0}\
\ \left[  {\ve}_{2}\, 
{\rm log} {\CZ} (a, {\ve}_{1} = {\ve} , {\ve}_{2}; {\qq})
\right]  = 
{\CW}^{\rm pert} (a; {\ve};
{\qq} ) 
 + {\CW}^{\rm inst} (a; {\ve};
{\qq} ) 
\label{eq:wform}
\end{equation}
Here the instanton part has an expansion in the
powers of $\qq$, 
\[ 
{\CW}^{\rm inst} (a; {\ve};
{\qq} )  = \sum_{k=1}^{\infty} \, {\qq}^{k}
{\CW}_{k}^{\rm inst}  (a; {\ve})
\] 
while the perturbative part has a tree level term, 
proportional to ${\rm log}({\qq})$ and
the one-loop term, which is ${\qq}$-independent.

It follows that in the limit ${\ve} \to 0$ 
the twisted superpotential ${\CW}(a, {\ve})$ 
behaves as:
\begin{equation}
{\CW}(a;  {\ve}; {\qq})= 
\frac{{\CF}(a; {\qq})}{\ve} + \ldots \
\label{eq:expan}
 \end{equation}
 with $\ldots$ denoting the regular in $\ve$ terms, 
 and 
 now the equations on the 
 supersymmetric vacua assumes the {\it Bethe}
 (\ref{eq:betbao}) form with the superpotential
  ${\CW}(a, \ve)$ (\ref{eq:wform}), (\ref{eq:expan}).

Our course is now pretty much set. 
We shall use the gauge theory knowledge
of the instanton partition functions ${\CZ}(a, {\ve}_1, {\ve}_2; 
{\qq})$ to extract, via (\ref{eq:wform}), 
the twisted superpotential
${\CW}(a, {\ve}; {\qq})$. 
The details of this procedure are
reviewed in the Section {\bf  6} where various ways of writing 
${\CZ}(a, {\ve}_1, {\ve}_2; {\qq})$ and extracting 
${\CW}(a, {\ve}; {\qq})$ are presented.  For the
purposes of the current discussion, the function 
${\CW}(a, {\ve}; {\qq})$ is known and the logic above gives the
desired equation for the supersymmetric vacua in the form of Bethe equations
(\ref{eq:betbao}).

By the philosophy of \cite{MNS, GS-one, GS-two, BAsh2008-one, BAsh2008-two} this superpotential serves as the Yang-Yang function of some
quantum integrable system. In the present context the parameter
${\ve}$ plays the r\^ole of the Planck constant, and can be tuned
to zero. In this limit we shall be able to use the quasiclassical asymptotics
(\ref{eq:expan}) to identify the classical integrable system,
whose quantization (in ${\ve})$ is the quantum integrable system
in question. 

To this end we need to remind the r\^ole of the prepotential 
${\CF}(a; {\qq})$
in the world of classical integrability \cite{KricheverBook}. 

{\bf Remark 1.} Note that in the derivation \cite{NOsw} 
of the 
Seiberg-Witten prepotential ${\CF}(a)$ from the direct instanton counting
 one evaluates the small ${\ve}_{1}, {\ve}_{2}
\to 0$ asymptotics of ${\CZ}(a ,  {\ve}_{1}, {\ve}_{2} ; {\qq} )$
by  a discrete version of the saddle point method, which connects
nicely the theory of instanton integrals to the theory of limit shapes
and random geometries. In our story
we need to go beyond that analysis,  see  Section {\bf 6}. Our results
suggest that the "quantum theory of the limit shape" is related
to the thermodynamic Bethe ansatz \cite{YangYang, TBAzam, Destri, TBAblz, TBAdor, TBAteschner}. 

{\bf Remark 2.} The simplest case of the 
$\Omega$-background is in two dimensions. 
The two dimensional $\Omega$-background is 
characterized by the single parameter ${\ve}$. The
partition function $I_{\al\be}({\ve})$ depends on two (discrete) parameters:
the choice $\be$ of the boundary condition at infinity, and the
choice of the twisted chiral ring operator ${\CO}_{\al}$
inserted at the origin. These partition functions
were studied in the context of the (equivariant) 
Gromov-Witten theory \cite{Givental}, where the ${\ve}$-dependence
comes from the coupling to the two dimensional
topological gravity. When the theory is deformed
in a way analogous to (\ref{eq:tdef}), the partition 
function becomes a matrix-valued function 
$I_{\al\be}({\bt}; {\ve})$ which solves the {\it quantum
differential equation}:
\begin{equation}
{\ve} \frac{\pa}{\pa t^{\gamma}} 
I_{\al\be}({\bt}; {\ve}) = 
C_{\al\gamma}^{\kappa}({\bt})
I_{\kappa\be}({\bt}; {\ve})
\label{eq:qdiff}
\end{equation}
where $C_{\al\be}^{\gamma}$ are the structure
constants of the twisted chiral ring. 

\newpage
\section{Integrable systems}

In this section we remind a few relevant notions in the theory of classical and quantum integrable systems and in particular explain the role of prepotential in classical algebraic integrable system. We introduce the
main examples, the periodic Toda, the elliptic Calogero-Moser
systems, their relativistic versions, and the Hitchin system. We also introduce (formally) the $t$-deformations of these systems, along the lines of \cite{LNS}.

\subsection{The classical story}

The classical Hamiltonian integrable system
is the collection $( {\CP}, {\om}, {\bH} )$,
where $\CP$ is a $2n$-dimensional smooth 
manifold endowed with the non-degenerate closed two-form $\om$ and a collection ${\bH} = (H_{1}, H_{2}, \ldots , H_{n})$ of (generically)
functionally independent functions ${\bH}:
{\CP} \longrightarrow {\bR}^{n}$, which mutually Poisson-commute: $\{ H_{i}, H_{j} \} = 0$, 
$i \neq j$. Here $\{ A, B \} = \left( {\om}^{-1} \right)^{\m\n} {\pa}_{\m} A {\pa}_{\n} B$. We can view $\CP$ as a Lagrangian fibration ${\bH}: {\CP} \to {\CU} \subset {\bR}^{n}$. 

The classical Liouville-Arnold theorem states that
if the common level set ${\bH}^{-1}(h)$ is compact, then it is diffeomorphic to the $n$-dimensional torus ${\bT}^{n}$. Moreover, if
${\bH}^{-1}(h)$ is compact for any $h$ in a neighborhood $\CU$ of a point $h_{0} \in {\bR}^{n}$, then ${\bH}^{-1}({\CU})$
is symplectomorphic to the neighborhood of a zero section in $T^{*}{\bT}^{n}$. 
One can then find the special Darboux coordinates $({\bI}, {\bf\varphi})$, ${\bI} = (I_{1}, I_{2}, \ldots , I_{n})$, ${\bf\varphi} =
({\varphi}_{1}, \ldots , {\varphi}_{n})$, called the {\it action-angle} variables, s.t.
the Hamiltonians $H_{i}$, $i=1,\ldots , n$, depend only on $\bI$, 
$H_{i}({\bI})$, while ${\varphi}_{i}$ are
the periodic angular coordinates on ${\bT}^{n}$ with the period $2\pi$.
Explicitly, the {\it action} variables are given by the periods
\begin{equation}
\label{eq:periodaction}
I_{i} = {1\over 2\pi}
\oint_{A_{i}} {\bp}{\dd}{\bq}
\end{equation}
of the one-form ${\bp}{\dd}{\bq} = {\dd}^{-1} {\om}$ (one can give a more invariant definition), over some ${\bZ}$-basis
in $H_{1}({\bT}^{n}, {\bZ})$.  

The notion of the classical {\it real} integrable system has an interesting {\it complex} analogue, sometimes known as the {\it algebraic
integrable system}. The data
$({\CP}, {\om}, {\bH})$ now
consists
of the complex manifold $\CP$,
the holomorphic non-degenerate
closed $(2,0)$ form $\om$, 
and the holomorphic map
${\bH}: {\CP} \to {\bC}^{n}$
whose fibers ${\CJ}_{h} = 
{\bH}^{-1}(h)$ are Lagrangian
polarized abelian varieties. The polarization is
a K\"ahler form ${\varpi}$, whose restriction on
each fiber is an integral class $[ {\varpi} ] \in 
H^{2} (J_{h}, {\bZ}) \cap H^{1,1}(J_{h})$. 
The image ${\CB} = {\bH} ({\CP})$ is an open
domain in ${\bC}^{n}$. It has a special 
K\"ahler
geometry, with the metric
\begin{equation}
\label{eq:dsdadad}
{\dd}s^{2} = 
\frac{1}{\pi}\sum_{i=1}^{n} {\rm Im}\, \left(
{\dd}a^{i} \otimes {\dd}{\bar a}_{D,i} \right)
\end{equation}
where the {\it special coordinates} $a^{i}, 
a_{D,i}$ are given by the periods:
\begin{equation}
\label{eq:aad}
a^{i} = {1\over 2\pi}
\oint_{A_{i}} {\bp}{\dd}{\bq}\, , \
a_{D,i} = {1\over 2\pi}
\oint_{B_{i}} {\bp}{\dd}{\bq}
\end{equation}
over the $A$ and $B$-cycles, which
are the Lagrangian (with respect to the
intersection form given by $[ \varpi ]$) subspaces in 
$H_{1}(J_{h},{\bZ})$. 
It follows that the two-form $\sum_{i}{\dd}a^{i} \wedge {\dd}a_{D,i}$ vanishes on $\CB$
thereby embedding the covering ${\CU}$ of the 
complement ${\CB} \backslash \Sigma$ to the discriminant $\Sigma \subset \CB$ of the singular fibers to the first cohomlogy $H^{1}(J_{h_{0}}, {\bC})$
of the fiber over some distinguished point $h_{0} \in {\CB}$, 
as a Lagrangian submanifold ${\CL}$. As such, it
comes with the function ${\CF}: {\CL} \to 
{\bC}$ which can be locally viewed as 
a function of $a^{i}$, such that
\begin{equation}
\label{eq:addf}
a_{D,i} = {{\pa}{\CF} \over {\pa}a^{i}}
\end{equation}
The comparison of the Eqs. (\ref{eq:aad}) and 
(\ref{eq:periodaction}) suggests both $a^{i}$ and $a_{D,i}$ are the {\it complex 
action variables}. Since the $2n$-dimensional  symplectic manifold has at most
$n$ functionally independent Poisson-commuting functions, there
ought to be a relation between $a^{i}$ and $a_{D,i}$'s. It is remarkable
that this relation has a potential function.  The action variables $a^{i}$ come with the corresponding
angle variables ${\phi}_{i} = {\al}_{i} + {\tau}_{ij}
{\be}^{j}$, ${\tau}_{ij} = {\pa}^{2}_{ij}{\CF}$, while $a_{D,i}$ correspond to ${\phi}^{D,i} = 
\left( {\tau}^{-1} \right)^{ij} {\phi}_{j}$:
\begin{equation}
{\om} = \sum_{i} {\dd}a^{i} \wedge {\dd}{\phi}_{i} = 
\sum_{i} {\dd}a_{D,i} \wedge {\dd}{\phi}^{D,i}
\label{eq:omg}
\end{equation}
Here
${\al}_{i}, {\be}^{i} \in {\bR}/2{\pi}{\bZ}$ are the
real angular coordinates on the Liouville torus.

\subsubsection{The ${\bt}$-deformation}

An algebraic classical integrable system $( {\CP}, {\om}, {\bH})$ can be deformed in the following way. Consider a family
$({\CP}_{\bt}, {\om}_{\bt}, {\bH}_{\bt})$ of complex symplectic manifolds with the Lagrangian fibration given by the ``Hamiltonians'' ${\bH}_{\bt}: {\CP}_{\bt} \to {\bC}^{n}$, over a (formal) multidimensional disk 
$\CDd$, parameterized by $\bt$. Then the variation of the symplectic form ${\om}_{\bt}$ in $\bt$ can be described by the equation:
\begin{equation}
\frac{\pa}{\pa t_{k}} {\om}_{\bt} = {\CL}_{V_{k}} {\om}^{\bR}
\label{eq:varom2}
\end{equation}
where
\[ {\om}^{\bR} = \sum_{i=1}^{n} {\dd}{\al}_{i} \wedge {\dd} {\be}^{i} + \sum_{i,j} {\rm Im} {\tau}_{ij} {\dd}a^{i} \wedge {\dd} {\bar a}^{j} \]
is the K\"ahler $(1,1)$-form, and $V_{k}$ is the holomorphic Hamiltonian 
(in the ${\om}_{\bt}$ symplectic structure) vector field corresponding
to the Hamiltonian $H_{k}$ of the original integrable system.

\subsubsection{Quantization}

The quantization of the classical integrable system is a (possibly discrete) family 
$({\CA}_{\ve}, {\CH}_{\ve}, {\hat\bH})$, 
of the associative algebras ${\CA}_{\ve}$, which deform the algebra of functions on the Poisson manifold $(X, {\om}^{-1})$, the (Hilbert) vector spaces ${\CH}_{\ve}$, with the action of ${\CA}_{\ve}$, and the
operators ${\hat\bH} = ( {\hat H}_{1}, \ldots , {\hat H}_{n})$, ${\hat H}_{i} \in {\CA}_{\ve}$, which mutually commute $[ {\hat H}_{i}, {\hat H}_{j} ] = 0$, $i \neq j$, and generate ${\CH}_{\ve}$ in the following sense: 
the common spectral problem
\begin{equation}
\label{eq:spec}
{\hat H}_{i} {\Psi} = {\CE}_{i} {\Psi}
\end{equation}
defines a basis in ${\CH}_{\ve}$.  Here $\ve$ has the meaning of the Planck constant.

The construction of the common eigenstates and the spectrum of the
operators ${\hat H}_{i}$  is a problem of the coordinate Bethe ansatz, quantum inverse scattering method (Algebraic Bethe ansatz 
\cite{Faddeevetc, FadTak, FaddeevLH}),
quantum separation of variables \cite{Zhenya}, Baxter equation \cite{Baxter}, the approach of \cite{fedya} to the spectral curve quantization
and various other versions of the Bethe ansatz.  

In this paper we take a different route - via the supersymmetric gauge theory, along the lines of  \cite{MNS, GS-one, GS-two, BAsh2008-one, BAsh2008-two}. The gauge theory allows to find the exact spectrum of (\ref{eq:spec}) which is the invariant  
of the choice of polarization used in the quantization procedure. The closest to our approach in 
the integrability literature seems to be that
 of \cite{fedya}. 

In our story the algebra ${\CA}_{\ve}$ 
is the deformation of the algebra of holomorphic functions
on $\CP$. We shall assume the existence of the global coordinates $p_{i}, x_{i}$ (we do not assume them to be
the globally defined holomorphic functions on $\CP$, but on some covering space). The algebra ${\CA}_{\ve}$ is generated by ${\hat p}_{i}, {\hat x}_{i}$, obeying 
\begin{equation}
\label{eq:pqc}
[{\hat p}_{i}, {\hat x}_{j} ] = {\ve} {\de}_{ij}\, ,
\end{equation}
${\CH}_{\ve}$ is defined as the space of appropriate
holomorphic functions of $x_{i}$, and the representation of 
${\CA}_{\ve}$ is given by:
\begin{equation}
\label{eq:repa}
{\hat x}_{i} = x_{i}, \ {\hat p}_{i} = {\ve} {{\pa} \over {\pa}x_{i}}
\end{equation}
The next question is the construction of the Hilbert space
where ${\CA}_{\ve}$ is represented. If ${\CP}$ were a cotangent
bundle to a complex manifold $M$, the algebra
${\CA}_{\ve}$ would be isomorphic to the algebra
of holomorphic differential operators on $M$. However,
it is rarely the case that there are interesting holomorphic
differential operators which are defined globally on $M$. 
Moreover, the naive complexification of the quantization of
the quantization of $T^{*}M_{\bR}$ (which produces
the differential operators on $M_{\bR}$ acting in 
the space of half-densities 
$L^{2}(M_{\bR}; K_{M_{\bR}}^{1/2})$)
produces the $K_{M}^{1/2}$ --
twisted differential operators. It may well happen that
the space of global sections of the $K_{M}^{1/2}$ --
twisted differential operators is non-zero yet the space
of global sections of $K_{M}^{1/2}$ where
these operators would have acted is empty. This is the situation
with the quantization of Hitchin system as discussed
in \cite{BeilinsonDrinfeld}. 

In our story the algebra ${\CA}_{\ve}$ {\it is}
the noncommutative deformation of the algebra
of holomorphic functions on $\CP$, yet it is
represented in the regular $L^2$-sections
of some line bundle on a real middle dimensional
submanifold ${\CP}_{\bR}$ of $\CP$. The choice
of ${\CP}_{\bR}$ is apparently made
by the boundary conditions in the gauge theory. 
We do not have a complete understanding of this
issue yet, but let us make an important:

{\bf Remark}: The equation (\ref{eq:main}) and the
discusion in {\bf Section 2} show that in our approach we 
quantize, for the type A model (defined below) 
the real submanifold which projects onto the locus  
${\rm Re} \left( {\pa {\CW}(a; {\bf t}; {\ve})} / {\pa a^{i}}\right) = 0$ in the base,
and cuts out a middle dimensional real torus in the Liouville fiber; 
for the type B model it projects onto the locus ${\rm Im} \left( a^{i}/{\ve} \right) = 0$. 

\subsubsection{Quantization and $\bt$-deformation}

The quantum integrable system can be deformed by making 
${\hat H}_{i}$ depend
on the additional parameters ${\bt} = (t_{1}, \ldots , t_{n})$, 
in parallel with the classical deformation  (\ref{eq:varom2}), so that they define a flat connection depending on a spectral parameter $\kappa$:
\begin{equation}
\label{eq:kpsp}
[ {\kappa}{{\pa}\over {\pa}t_{i}} - {\hat H}_{i}({\bt}) , \  {\kappa}{{\pa}\over {\pa}t_{j}} - {\hat H}_{j}({\bt}) ] = 0
\end{equation}
for all $i, j = 1, \ldots , n$. This deformation, when applied
to quantum Hitchin system,  is related
to the KZB connection in the WZW conformal field theory. The gauge theoretic meaning of the $\kappa$-parameter is ${\ve}_2$ of the general
$\Omega$-background.

\newpage
\section{Examples}

We now proceed with explicit examples.

\subsection{The periodic Toda chain}

\subsubsection{The classical system}
The periodic Toda chain is the system of $N$ particles
$x_{1}, \ldots , x_{N}$
on the real line interacting with the potential:
\begin{equation}
\label{eq:pottoda}
U (x_{1}, \ldots , x_{N}) = {\Lambda}^{2} \left( \sum_{i=1}^{N-1}
e^{x_{i}-x_{i+1}} + e^{x_{N}-x_{1}} \right)
\end{equation}
The phase space of this model is
${\CP}_{\bR} = T^{*}{\bR}^{N}$, with the coordinates $(p_{i}, x_{i})_{i=1}^{N}$, where $p_{i}, x_{i} \in {\bR}$, the symplectic
form ${\om} = \sum_{i=1}^{N} {\dd}p_{i} \wedge {\dd}x_{i}$, and the Hamiltonians
\begin{eqnarray}
H_{1} & = & \sum_{i} p_{i} \\
H_{2} & = & {1\over 2} \sum_{i} p_{i}^{2}  + U (x_{1}, \ldots , x_{N}) \label{eq:hamtoda}\\
& \ldots &  \\ 
H_{k} & = & {1\over k!} \sum_{i} p_{i}^{k} + \ldots \\
\end{eqnarray}
The complexified Toda chain has
 the phase space
${\CP} = 
T^{*} \left( {\bC}^{\times} \right)^{N}$, with the
coordinates $(p_{i}, x_{i})_{i=1}^{N}$ where
$p_{i} \in {\bC}, x_{i} \in {\bC}/(2\pi i){\bZ}$. 
   
To describe this model as the algebraic integrable
system we introduce the Lax operator
\begin{equation}
\label{eq:laxtod}
{\Phi}(z) = \begin{pmatrix} 
p_{1} & {\Lambda}^{2} e^{x_{1}-x_{2}} & 0 & \ldots & \ldots & 
e^{-z} \\
1 & p_{2} & {\Lambda}^{2} e^{x_{2}-x_{3}} & \ldots & \ldots & 0\\
0 & 1 & p_{3} & {\Lambda}^{2} e^{x_{3}-x_{4}} & \ldots & 0 \\
0 &\ldots &  \ldots & \ldots & \ldots &  0 \\
0 &\ldots &  \ldots & \ldots & \ldots &  0 \\
0 &\ldots &  \ldots & \ldots & p_{N-1} &  {\Lambda}^{2} e^{x_{N-1}-x_{N}} \\
{\Lambda}^{2} e^{x_{N}-x_{1}} e^{z} & 0 & \ldots
& 0 & 1 & p_{N} \\
\end{pmatrix}
\end{equation}
and define the Hamiltonians $h_{1}, \ldots , h_{N}$
as the coefficients of the characteristic polynomial:
\begin{equation}
\label{eq:detcur}
{\mathrm{Det}} \left(  x-{\Phi}(z)\right) = 
-{\Lambda}^{2N} e^{z} - e^{-z} + x^{N} + h_{1} x^{N-1} + h_{2} x^{N-2} + \ldots + h_{N}
\end{equation} 
 The Hamiltonians $h_{1}, h_{2}$ are then given by
 \[
 h_{1} = - \sum_{i=1}^{N} p_{i}
 \]
 and
\begin{equation}
\label{eq:htwo}
h_{2} = -\sum_{i < j} p_{i} p_{j} + U (x_{1}, \ldots , x_{N})
\end{equation}
which differs from $H_{2}$ defined in (\ref{eq:hamtoda})
by the term $\frac{1}{2}h_{1}^{2}$. 

Define the {\it spectral curve} ${\CC}_{h} \subset
{\bC} \times {\bC}^{\times}$, $x \in {\bC}, z \in 
{\bC}/2\pi i$, as the zero locus
of the characteristic polynomial (\ref{eq:detcur}). 
For each value 
$h = (H_{1}, H_{2}, \ldots , H_{N})$ this is a
curve which is a genus $N-1$ hyperelliptic
curve with two points where $x = \infty$ deleted.  The fiber ${\bH}^{-1}(h)$ 
 is given by the product
${\bC} \times {\CJ}_{h}$. The 
${\bC}$-factor corresponds to the center-of-mass
mode $\sum_{i} x_{i}$, the compact
factor ${\CJ}_{h} = {\rm Jac}(\overline{{\CC}_{h}})$ is the Jacobian
of the compactified curve ${\CC}_{h}$. 

The complex action variables $a^{i}$, $a_{D,i}$
can be computed as the periods of the differential
\begin{equation}
\label{eq:swd}
{\la} = {1\over 2\pi} x {\dd}z
\end{equation}

\subsubsection{The gauge theory}
The gauge theory significance of the periodic Toda chain is that
the potential ${\CF}(a)$ in the Eq. (\ref{eq:addf}) defined using the family
of spectral curves  (\ref{eq:detcur}) and the differential (\ref{eq:swd})
coincides with the prepotential of 
 the low energy effective Lagrangian of the  pure $\CN=2$ gauge theory
 with the gauge group $U(N)$.

\subsubsection{The quantum system}

The quantization of the periodic Toda chain is achieved by promoting
$h_k$'s defined by (\ref{eq:detcur}) to the differential operators acting on
functions of $(x_{1}, \ldots , x_{N})$, $p_{i} = {\ve} {\pa}_{x_{i}}$. 
It is possible to show that the potential normal ordering ambiguities do not 
arise in this case. One is looking for the eigenfunctions of the form:
\begin{equation}
{\Psi}(x_{1}, \ldots , x_{N}) = e^{ N k {\bar x}}\,
{\psi}_{k} ( x_{1} - {\bar x}, x_{2} - {\bar x}, \ldots , x_{N}- {\bar x})
\label{eq:com}
\end{equation}
where \[
{\bar x} = \frac{1}{N} \sum_{i=1}^N x_{i} \]
${\psi}_{k} \in L^2 ( {\bR}^{N-1})$. 
When ${\ve} = - i{\hbar}$ and ${\Lambda}, {\hbar} \in {\bR}$
this problem has a discrete real spectrum (for fixed $k$). We are after the effective characterization of this spectrum. It turns out that this model
allows a complex analytic continuation in the parameters ${\ve}$, 
${\Lambda}$, so that the spectrum remains discrete, yet
in general complex. We shall call this spectral problem the type 
A quantum periodic Toda.

 The quantum periodic Toda admits another,
somewhat unconventional (for $N > 2$) formulation, which also
leads to the discrete yet complex spectrum. In this formulation we
make the differential operators ${\hat H}_{k}$ act on functions
${\Psi} (x_{1}, \ldots, x_{N}) $ which are $2\pi i$-periodic,
and non-singular for some fixed value of ${\rm Re} x_{1}, \ldots
, {\rm Re} x_{N}$. We shall call this spectral problem the type B
quantum periodic Toda. 

Note that for $N =2$ case the type B periodic Toda is equivalent
to finding the (quasi)-periodic solutions of the canonical
Mathieu's differential equation, while the type A model
corresponds to the $L^2$ solutions of 
Mathieu's modified differential equation.

{\bf Remark.} The quasi-periodic solutions of the differential equations
are parametrized by the so-called Bloch-Floquet
multipliers, which in our correspondence come from
the two-dimensional theta angle ${\vt}$, which
might come from the $B$-field in four dimensions,
or from the peculiar deformation of the
four dimensional tree level prepotential $\propto
{\ve} {\vt}\, {\tr}{\Phi}$, in the 
$\Omega$-background ${\bR}^2_{\ve}$.  
Of course we can
set ${\vt} = 0$ and discuss the periodic wavefunctions. 
This remark applies to all the many-body systems.

\subsection{Elliptic Calogero-Moser system}

\subsubsection{The classical system}

The elliptic Calogero-Moser system (eCM) is the system of $N$ particles 
$x_{1}, x_{2}, \ldots , x_{N}$ on the circle of circumference $\be$, i.e. 
$x_{i} \sim x_{i} + {\be}$, which interact with the pair-wise potential
\begin{equation}
 U(x_{1}, x_{2}, \ldots , x_{N})  =  m^{2}
\sum_{i <  j} u (x_{i}- x_{j})\, , \label{eq:pairpot}\end{equation}
\[
  u (x)  = C({\be}) + 
 \sum_{k\in {\bZ}} {1\over {\rm sinh}^{2}( x + k {\be} ) } 
 = - {\pa}^2_{x}
 \,\, {\rm log}\, {\Theta}(x) \]
 where $\Theta$ is the odd theta function
on the elliptic curve $E_{\tau}$ with the modular parameter
${\tau} = {i  {\be} \over \pi}$,
\[
{\Theta}(x) = - \sum_{k \in {\bZ}+\frac{1}{2}}
(-1)^k {\qq}^{\frac{k^2}{2}} e^{2 k x }, \qquad
{\qq} = {\exp} \, 2\pi i \tau
\]
and $C({\be})$ is some constant which depends on $\be$.

 Again, there exists a Lax representation \cite{KricheverCM} of the eCM
system:
\begin{equation}
\label{eq:laxecm}
{\Phi}_{ij}(z) = p_{i} {\de}_{ij} + m
{{\Theta} (z + x_{i} - x_{j}) {\Theta}^{\prime}(0) \over
{\Theta}(x_{i} - x_{j} ) {\Theta}(z)} (1 - {\de}_{ij}) 
\end{equation}

In the limit ${\be} \to \infty$, $m \to \infty$, such that 
${\Lambda}^{2N} = m^{2N} {\qq}$
is kept finite, the eCM system becomes the periodic
Toda chain \cite{Inosemtsev}, 
where
\[ x_{i}^{\rm eCM} = \frac{i}{N}\, {\be} +
  x_{i}^{\rm pToda} \]

\subsubsection{The gauge theory}

The spectral curve ${\rm Det}( {\Phi}(z) - x) = 0$, which is an $N$-sheeted
ramified cover of the elliptic curve $E_{\tau}$ where $z$ lives, is the
Seiberg-Witten curve of a remarkable four dimensional ${\CN}=2$
theory. This is an $SU(N)$ gauge theory with a massive
adjoint hypermultiplet. The parameter $m$ in (\ref{eq:laxecm})
is the mass of the adjoint hypermultiplet, the complex structure
of the curve $E_{\tau}$
is determined by the complexified bare gauge coupling of the ultraviolet
theory (which is in fact the superconformal  ${\CN}=4$ super-Yang-Miills). 
We shall call this theory the ${\CN}=2^{*}$ theory.

\subsubsection{The quantum system}

\begin{equation}
\label{eq:hamtwo}
{\hat H}_{2} = {{\ve}^{2}\over 2}
\sum_{i=1}^{N} {{\pa}^{2} \over {{\pa}x_{i}^{2}}} - m ( m +{\ve} ) \sum_{i< j}
{\wp} ( x_{i} - x_{j} ) 
\end{equation}
When ${\ve} = - i {\hbar}$, $m = - {\nu}{\ve}$, and
${\hbar}, {\nu}, {\be} \in {\bR}_{+}$ the spectral problem of the
operator (\ref{eq:hamtwo}) is well-known: one is looking for the
(quasi-)periodic (in ${\be}$) symmetric functions ${\Psi}(x_{1}, \ldots , 
x_{N})$, which are non-singular in the fundamental domain, and
vanish at the diagonals:
\begin{equation}
{\Psi}(x_{1}, \ldots , x_{N} ) \sim ( x_{i} - x_{j})^{\nu}
\label{eq:diag}
\end{equation}
It follows from our results that this problem has an analytic continuation
where ${\ve}, m, {\be}$ all become complex. The monodromy of this continuation
is an extremely interesting problem which we shall not be able to cover
in this exposition. 

Just like in the case of the periodic Toda chain one may consider
various spectral problems for the Hamiltonian ${\hat H}_2$
and its higher order counterparts. The pleasant bonus of having
an elliptic potential is the similarity of these problems. The type A
quantum elliptic Calogero-Moser is, therefore, a problem which we
just described: 
finding the ${\be}$-periodic $L^{2}$ (on the subspace
with fixed ${\rm Im}(x_{i}/{\be})$)
wavefunctions, with the
(\ref{eq:diag}) behavior. The type B problem is a problem of
finding the ${\pi} i$- periodic $L^2$ wavefunctions (on the subspace
with fixed ${\rm Re}(x_{i})$). The $SL_{2}({\bZ})$
transformation ${\tau} \to -\frac{1}{\tau}$ maps the type A problem
to the type B problem and vice versa.

Superficially, however, one may worry that the type A and the type
B problems are quite different in nature. In the limit ${\be} \to \infty$
the type A problem becomes that of the hyperbolic Sutherland
model, which has a continuous spectrum. In the same limit
the type B problem becomes that of the trigonometric
Sutherland, which is well-studied, has a discrete spectrum, and
Jack polynomials (up to a ground state factor) as the eigenfunctions. 
Our approach connects all these models.

In the $N=2$ case we are solving the celebrated Lame's differential
equation \cite{Lame}.

\subsection{Hitchin system}

The previous systems are the degenerate examples
of the so-called Hitchin integrable system.
Its phase space ${\CP}$ is the partial resolution
of the cotangent bundle to the moduli space
$\CM$ of semistable holomorphic bundles on
a complex curve $C$. 
More precisely, the phase space is the space
of pairs $( {\nabla}_{\pb}, {\Phi} = {\Phi}_{z}{\dd}z )$, where
${\nabla}_{\pb} = {\pb} + {\bar A}$, ${\bar A} = A_{\zb} {\dd}{\zb}$, is the $(0,1)$-part
of a connection on a vector bundle $E$ over $C$, and ${\Phi}_{z}$ is the
Higgs field, a $(1,0)$-form valued in the endomorphisms of the $E$. The vector bundle
$E$ gets a holomorphic bundle structure by
declaring the solutions to
${\nabla}_{\zb} {\chi} = 0$ to be the holomorphic
sections. 
The Higgs field $\Phi_{z}$ must obey:
\begin{equation}
\label{eq:hitcheq}
{\nabla}_{\pb}{\Phi} = 0, \, i.e.\
{\pb}_{\zb}\Phi_{z} + [ A_{\zb}, {\Phi}_{z}] = 0
\end{equation}
Divide by the complex gauge transformations,
or make a symplectic quotient with respect to the
compact gauge transformations. In the latter case one imposes the real moment map condition:
\begin{equation}
\label{eq:rmom}
F_{A} + [ {\Phi}, {\bar\Phi} ] = 0
\end{equation} 
The holomorphic symplectic structure is induced from
\begin{equation}
\label{eq:symc}
{\om} = \int_{C} {\tr} \, {\de}{\Phi}\wedge {\de} {\bar A}
\end{equation}
The Hamiltonians are obtained as follows:
consider the $(j,0)$-differentials on $C$, given by
${\tr}{\Phi}^{j}$. Due to (\ref{eq:hitcheq}) these are holomorphic $j$-differentials, and (for $j>1$) there
are $(2j-1)(g-1)$ linearly independent (over $\bC$) such differentials. The base ${\CB}$ of the corresponding integrable system is the vector space:
\begin{equation}
\label{eq:bshith}
{\CB} = \oplus_{j=1}^{N} H^{0} \left( C, K_{C}^{\otimes \, j} \right) \approx
{\bC}^{1 + N^{2}(g-1)}
\end{equation}
while the fiber over a point $h \in {\CB}$ is the Jacobian of the spectral curve:
\begin{equation}
\label{eq:speccr}
{\CC} \subset T^{*}C\, : \qquad
0 = {\Det} \left( {\Phi} - x \right) 
\end{equation}
(where we view $x$ as the canonical Liouville one-form
on $T^{*}C$). 
The homology class $[{\CC}]$ spectral curve  
is equal to $N [ C ]$. It follows that the self-intersection
number $2g({\CC}) -2 = {\CC}.{\CC} = N^{2} C.C = N^{2} (2g-2)$, therefore the genus 
\begin{equation}
\label{eq:genspec}
g({\CC}) = 1 + N^{2} (g-1)
\end{equation}
is equal to the dimension of $\CB$. 
The polarization comes from the Kahler form:
\begin{equation}
\label{eq:kahfrm}
{\varpi} = \int_{C} \, {\tr} \left( {\de}A \wedge {\de}{\bar A} + {\de}{\Phi} \wedge {\de} {\bar\Phi} \right)
\end{equation}
Let us now explain that the systems we considered so far are the degenerate
cases of Hitchin system. Imagine we study a Hitchin system for the group 
$U(N)$ on a curve of genus two. Now let us degenerate the curve, in such a way
that it becomes a union of two elliptic curves, connected by a long neck. 
Then the equations (\ref{eq:hitcheq}) can be solved on both elliptic curves
independently, the only memory of the original curve being a boundary condition at the puncture representing the neck. The limiting Hitchin system
would actually split as a union of invariant submanifolds, labelled
by the coadjoint orbits ${\CO}$ of the complexified gauge group, attached to the
puncture (as a hyperk\"ahler manifold it fibers over 
$({\bt} \otimes {\bR}^3)/W$) \cite{NekHol}. The limiting equations 
(\ref{eq:hitcheq}) look like:
\begin{equation}
{\pb}_{\zb}\Phi_{z} + [ A_{\zb}, {\Phi}_{z}] = J {\de}^{(2)}(z,{\zb})
\label{eq:hitcheqsrc}
\end{equation}
where $J$ represents a (complex) moment map of the group action on 
a coadjoint orbit $\CO$. Now let us consider a very special vase, where
the orbit ${\CO}$ is diffeomorphic to $T^* {\bC\bP}^{N-1}$. 
The corresponding moment map $J$ is the $N \times N$ complex
traceless matrix of which $N-1$ eigenvalues coincide,
\[ 
J = m\, {\rm diag} \left( N-1 , -1 , \ldots , -1 \right) \ .
\] 
By solving the equation
(\ref{eq:hitcheqsrc}) one gets (up to an irrelevant gauge transformation)
the Lax operator (\ref{eq:laxecm}) in the gauge where 
\[
A_{\zb} = {\rm diag} \left( x_{1}, \ldots , x_{N} \right)
\]
is a constant diagonal matrix \cite{NekGor}. 

\subsubsection{The quantum system and the gauge theory}

We do not have much to say about these two topics in the case
of  Hitchin system for compact Riemann surface. The supersymmetric field 
theory realization of this system involves the six dimensional
$(2,0)$ theory compactified on the Riemann surface $C$. 
Using the results of \cite{Gaiotto} one may hope
to formulate the result as the four dimensional ${\CN}=2$
supersymmetric gauge theory, at least for the curves $C$ which are
close to the degeneration locus. This gauge theory  can be further
subject to the $\Omega$-background. As a result
one would get a quantization of Hitchin system, together
with an expression for its Yang-Yang function. It would be
interesting to understand the relation
of our approach to the constructions of \cite{WittenKapustin}.

\subsection{Relativistic systems}

So far we considered the systems which have a Galilean symmetry group, the space translations are generated
by ${\hat H}_{1}$ while the time translations are generated
by ${\hat H}_{2}$, and there is a generator of the
the Galilean boost ${\hat S}$, such that 
$[ {\hat S}, {\hat H}_{2}] = {\hat H}_{1}$. 
The point is that the Hamiltonians 
${\hat H}_{1}, {\hat H}_{2}, \ldots , {\hat H}_{N}$ are polynomials in the momenta ${\hat p}_{1}, \ldots , {\hat p}_{N}$. 

There exists \cite{RS-i, RS-ii} a one-parametric deformation
of the eCM integrable system, whose Hamiltonians
are trigonometric (hyperbolic) functions of
the momenta. In particular, the Hamiltonians
${\hat P} =  {\hat H}_{1}, {\hat E} = {\hat H}_{2}$ become
\begin{equation}
{\hat P} =
\sum_{i=1}^{N} {\rm sinh} ({\be}p_{i}) 
f_{i}(x)\, , 
\label{eq:hath}
\end{equation}
\[
 {\hat E} =
\sum_{i=1}^{N} {\rm cosh} ({\be}p_{i}) 
f_{i}(x)\, , \]
\[  f_{i}(x) = \prod_{j \neq i} 
\sqrt{1  -  {{\wp} (x_{i}-x_{j}) \over {\wp}({\be}{m})}} 
\]
The quantization of the system (\ref{eq:hath}) produces a family
of commuting difference operators. These operators, in the trigonometric
limit, reduce (after a similarity transformation), to the famous Macdonald
difference operators \cite{Macdo}, whose eigenfunctions are the so-called 
Macdonald polynomials. In the context of a physical problem, the polynomial
eigenfunctions correspond to the compact model, where the particles
$x_{1}, \ldots , x_{N}$ live on a circle.  The problem of interest
for us reduces, in the trigonometric limit, to the hyperbolic
system where the particles live on a real line, and the spectrum, as in the
non-relativistic case, becomes continuous. As far as we know, 
there is no satisfactory treatment of this model in the literature.
Therefore our results, presented below, might be of
additional interest for the community working in the domain of 
harmonic analysis. 

In our analysis, the relativistic systems (\ref{eq:hath}) correspond to the
five dimensional gauge theory compactified on a circle 
\cite{NekFD}, 
subject further to the $\Omega$-background in ${\bR}^{2}_{\ve}$. 

\newpage
\section{Superpotential/Yang-Yang function ${\CW}(a, {\ve}; {\qq})$}

In this section we show how ${\CW}(a, {\ve}; {\qq})$ is computed and give 
several representations for it. The ${\CW}(a, {\ve}; {\qq})$ from this section needs to be inserted in (\ref{eq:betbao}), (\ref{eq:cecho}) in order to write the explicit expression for the spectrum.

\subsection{Thermodynamic Bethe Ansatz}

The perturbative part of $\CW$ is written using the special
function ${\varpi}_{\ve}(x)$. It obeys
\[
{d\over dx} {\varpi}_{\ve}(x) = {\rm log} \, {\Gamma} \left( 1 + {x\over {\ve}} \right) \ .
\]
The instanton part of $\CW$ is given by the critical value of the following
functional:
\[
{\CW}^{\rm inst}(a ; {\qq}) = \]
\begin{equation}
\quad {\rm Crit}_{{\rho}, {\varphi}} \
\frac{1}{2}\int_{{\CC}\times {\CC}} 
{\rho}(x) {\rho}(y) G ( x - y) + \int_{\CC}\left[  {\rho}(x){\varphi}(x) +
{\rm Li}_{2} \left( {\qq} Q(x) e^{-{\varphi}(x)} \right) \right]
\label{eq:tbai}
\end{equation}
Equivalently:
\begin{equation}
{\CW}^{\rm inst}(a ; {\qq}) = \int_{\CC}\left[  -\frac{1}{2}
{\varphi}(x) {\rm log} \left( 1 - {\qq} Q(x) e^{-{\varphi}(x)} \right) +
{\rm Li}_{2} \left( {\qq} Q(x) e^{-{\varphi}(x)} \right) \right]
\label{eq:tbaii}
\end{equation}
where ${\varphi}(x)$ solves a TBA-like equation:
\begin{equation}
{\varphi}(x) = \int_{\CC} G( x- y) \,
{\rm log} \left( 1 - {\qq} Q(y) e^{-{\varphi}(y)} \right) 
\label{eq:tbaiii}
\end{equation}
The solution is determined by the choice of the contour ${\CC}$,
and the functions $G(x)$, $Q(x)$. Given this data, it is straightforward to solve
(\ref{eq:tbaiii}) recursively:
\begin{equation}
{\varphi}(x) = \sum_{k=1}^{\infty} {\qq}^k {\varphi}_{k}(x),\
 {\rho}(x) = \sum_{k=1}^{\infty} {\qq}^k {\rho}_{k}(x)
\label{eq:vrphi}
\end{equation}
\subsection{The examples}

We now give the expressions for the functions
$Q(x), G(x)$ and the contour $\CC$, for our examples.

\subsubsection{Periodic Toda}

For the periodic Toda chain one has a simple formalism with
\[ Q(x) = \frac{1}{P(x)P(x+{\ve})}\, , \, G(x) = {d\over dx}\,  {\rm log}
  {( x - {\ve}) \over( x + {\ve})  } \]
\begin{equation}
{\CW}^{\rm pert} = {1\over 2\ve} {\rm log} \left( {{\Lambda} \over \ve} \right) \sum_{n=1}^{N} a_{n}^2 + \sum_{l,n=1}^{N} \,  {\varpi}_{\ve}( a_{l} - a_{n})
\label{eq:pertoda}
\end{equation}
and the contour ${\CC}$ that goes around the points
$a_{l} + k {\ve}$, $k \geq 0$. It can be deformed to go along
the real line, when $a_{l} \in {\bR} + \frac{\ve}{2}$, and
${\rm Im}{\ve} >0$.

In the perturbative limit the Bethe equation derived from the superpotential 
(\ref{eq:pertoda})  assumes the form:
\begin{equation}
\left( \frac{\Lambda}{\ve} \right)^{2 N a_{i} \over \ve} = \prod_{j\neq i}\, - 
\frac{{\Gamma}\left( 1 + \frac{a_{i} - a_{j}}{\ve} \right)}{{\Gamma}\left( 1 - \frac{a_{i} - a_{j}}{\ve} \right)}
\label{eq:smattoda}
\end{equation}
which has the form of the ordinary Bethe ansatz equations for a system
of interacting
particles $(x_{1}, \ldots , x_{N})$ with the factorizable
$S$-matrix, which coincides with that of the open Toda chain. The two-body $S$-matrix is that of the Liouville quantum mechanics.

\subsubsection{The elliptic Calogero-Moser system}

Let $P(x) = \prod_{l=1}^{N} ( x - a_{l} )$, and
\begin{equation}
Q(x) = \frac{P(x-m)P(x + m + {\ve})}{P(x) P(x+ {\ve})}\, , 
\label{eq:pqfun}
\end{equation}
\[ 
G(x) = {d\over dx}\,  {\rm log}  {(x + m  + {\ve}) ( x - m ) ( x - {\ve}) \over ( x - m - {\ve}) ( x + m ) ( x + {\ve})  } \]
Then:
${\CW}(a, {\ve}; {\qq} ) = {\CW}^{\rm pert} + {\CW}^{\rm inst}$,
\begin{equation}
{\CW}^{\rm pert} =  {1\over 2\ve} {\tau} \sum_{n=1}^{N} a_{n}^2 + \sum_{l,n =1}^{N} \, \left( {\varpi}_{\ve}( a_{l} - a_{n})
- {\varpi}_{\ve}( a_{l} - a_{n} - m - {\ve} ) \right)
\label{eq:pert}
\end{equation}
and ${\CW}^{\rm inst}$ is given by (\ref{eq:tbaii}) with 
the contour $\CC$ in the complex plane that
comes from infinity, goes around the points $a_{l} + k {\ve}$, $l = 1, \ldots, N$, $k = 0, 1,2, \ldots $, and goes back to infinity. It separates these points
and the points $a_{l} + l m + k {\ve}$, $l \in {\bZ}$, $k = - 1, -2, \ldots $.  
\begin{figure}
\psfig{file=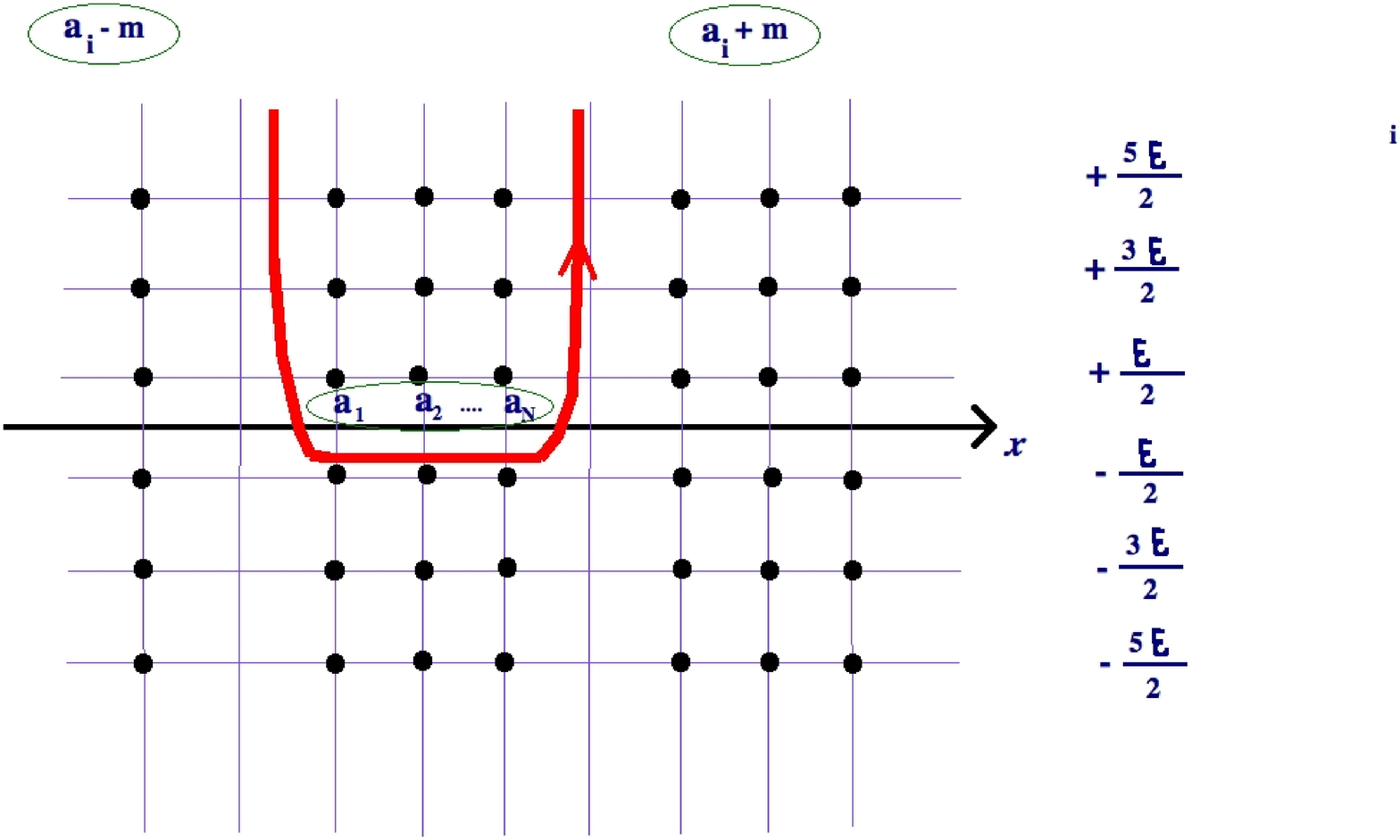, width=5in}
\caption{The contour ${\CC}$}
\label{aba:fig1}
\end{figure}
The perturbative Bethe equations have the form similar to (\ref{eq:smattoda}):
\begin{equation}
{\qq}^{a_{i} \over \ve} = \prod_{j\neq i}\,  
\frac{{\Gamma}\left( \frac{a_{i} - a_{j}}{\ve} \right)}{{\Gamma}\left( \frac{a_{i} - a_{j}}{\ve} \right)}
\frac{{\Gamma}\left( \frac{-m - a_{i} + a_{j}}{\ve} \right)}{{\Gamma}\left( \frac{-m + a_{i} - a_{j}}{\ve} \right)}
\label{eq:smatecm}
\end{equation}
which corresponds to the system of $N$ interacting particles on the circle of the size $\propto {\rm log}{\qq}$, 
 with the factorizable $S$-matrix of the hyperbolic Sutherland model.

\subsubsection{Ruijsenaars-Schneider model}

Let us denote: $q = e^{-{\be}{\ve}}, t = e^{- {\be}m}, z = 
e^{- {\be}x}, w_{l} = e^{{\be}a_{l}}$.
The perturbative contribution is:
\[
{\CW}^{\rm pert} = {{\pi i \tau}\over 2\ve}  \sum_{n=1}^{N} a_{n}^2 + \sum_{l,n =1}^{N} \, \left( {\Pi}_{q}( w_{l}/w_{n})
- {\Pi}_{q}( t  w_{l}/w_{n} ) \right) \]
where
\[ 
{\Pi}_{q} (w) = \frac{1}{\be} \sum_{n=1}^{\infty} \frac{w^n}{n^2 (1-q^{n}) } \sim \sum_{n \in {\bZ}} {\varpi}_{\ve} \left( \frac{{\rm log}(w) + 2\pi i n}{\be} 
 \right)  \]
The $P,Q,G$ functions are given by:
\begin{equation}
P(x) = \prod_{l=1}^N ( 1 - w_{l} z ), \ Q(x) = 
\frac{{P}( z t^{-1}) {P}( q t  z ) }{{P}(z) {P}(q z) }
\label{eq:trigpq}
\end{equation}
\[
G(x) = z {d\over dz}\, {\rm log} \left(  \frac{z^2 ( z - tq) 
( 1 - t z) 
( 1 - q z ) }{(z - t) ( z - q ) ( 1 - q t z )} \right)
\] 
We assume $|q|, |t|, |w_{l}|  < 1$, 
and the contour ${\CC}$ is the unit circle $|z|=1$. 

\subsubsection{The spectrum of observables}

The spectrum
of the type A quantum system is given by the
solutions of the equations (\ref{eq:betbao})
with ${\CW}(a, {\ve})$ given by (\ref{eq:tbai}). 
Let us denote the extremum of the functional
(\ref{eq:tbai}) by ${\rho}_{A}(x), {\varphi}_{A}(x)$. 
The spectrum
of the type B quantum system is given by the solutions
of the "dual" equations
\begin{equation}
a_{i} = {\ve} n_{i} \, , \qquad n_{i} \in {\bZ}
\label{eq:magn}
\end{equation}
With these $a_i$'s one can again define the functional
(\ref{eq:tbai}) and study its extremum ${\rho}_{B}(x),
{\varphi}_{B}(x)$. 

{\it The eigenvalues of the quantized
Hamiltonians 
\[
H_{k} = \int {\dd}z{\dd}{\zb}\, 
{\tr} {\Phi}(z)^{k} = \sum_{l=1}^{N} p_{l}^{k} + \ldots :
\]
are given respectively, for the type $A$ and type $B$ models:
\begin{equation}
{\CE}_{k} = \sum_{l=1}^{N} a_{l}^{k} + 
k \int_{\CC}\, {\dd}x \left( ( x+ {\ve})^{k-1} - x^{k-1} 
\right)  {\rho}_{A,B}(x) \, , 
\label{eq:energy}
\end{equation}
In particular:
\begin{equation}
{\CE}_2 = {\ve} {\qq} {d\over d{\qq}} {\CW}(a ; {\qq})
\label{eq:entwo}
\end{equation}
}

\subsubsection{On the relation between the type A and the type B models}

The ${\CN}=2^{*}$ gauge theory becomes the ${\CN}=4$ super-Yang-Mills
theory in the ultraviolet. The latter has the celebrated Montonen-Olive $SL_{2}({\bZ})$ symmetry. In particular, the $S$-duality transformation
maps the gauge theory with the gauge group $G$ and the coupling ${\tau}$
to the gauge theory with the gauge group $^{L}G$ and the coupling $-\frac{1}{h\tau}$, for some integer $h = 1, 2,3$. 
When the theory is perturbed by the mass term, the $S$-duality symmetry
still acts. We claim it maps the type A model of the elliptic Calogero-Moser
system with the modular parameter ${\tau}$ to the type B model
with the modular parameter $-\frac{1}{\tau}$. The special coordinates $a_{i}$
map to ${\tau} a_{i}$. The modularity
of the effective twisted superpotential is clearly supported by the expansion (\ref{eq:cwexp}).

\subsubsection{More details and the origin of integral equation}

Here we give more details in regard to the origin of claims from previous sub-sections. We do it for the case of the ${\CN}=2^{*}$ theory (the relativistic case
is studied analogously, along the lines of \cite{NekFD, ABCD}). One starts with the contour integral representation \cite{MNS, MNSbound, LNStest, Neksw} for the instanton partition function in the general $\Omega$-background:
\begin{equation}
{\CZ}^{\rm inst}({\ba}; {\qq}, m, {\ve}_{1}, {\ve}_{2}) =  \label{eq:instpzc} \end{equation}
\[
 \qquad \sum_{k=0}^{\infty} {{\qq}^{k}\over k!} \int_{{\bR}^{k}}
\prod_{1 \leq I < J \leq k}
{\CDd} ({\phi}_{IJ}) \prod_{I=1}^{k} \, Q ( {\phi}_{I}) \, {{\ve} \, ( m + {\ve}_{1} ) (m + {\ve}_{2} ) \over {\ve}_{1} {\ve}_{2}\, m (m + {\ve}) } { {\rm d}{\phi}_{I} \over 2\pi i} \, , \]
where ${\phi}_{IJ} = {\phi}_{I} - {\phi}_{J}$, 
\[
{\CDd}(x) = \frac{x^{2} ( x^{2} - {\ve}^{2}) ( x^{2} - (m+{\ve}_{1})^{2} ) ( x^{2} - ( m + {\ve}_{2})^{2})}{( x^{2}  - {\ve}_{1}^{2} ) ( x^{2} - {\ve}_{2}^{2}) ( x^{2} - m^{2} ) ( x^{2} - ( m + {\ve})^{2})} , \]
 and $Q(x)$ was introduced in (\ref{eq:pqfun}). 
Next use the observation, reported earlier in \cite{NekAmsterdam, MN}, that (\ref{eq:instpzc}) is a partition function of a one-dimensional non-ideal gas of particles ${\phi}_{1}, \ldots , {\phi}_{k}$
subject to the external potential 
\begin{equation}
U_{\rm ext} (x) = - {\rm log} \left( Q(x) \frac{(m + {\ve}_{1})(m + {\ve}_{2} )  {\ve}} {{\ve}_{1}{\ve}_{2} m ( m + {\ve})} \right)
\label{extpot}
\end{equation}
and a pair-wise interaction potential
\begin{equation}
V_{\rm int} (x) = - {\rm log} \left(  {\CDd}(x) \right) 
\label{intpot}
\end{equation}
The free energy of this gaz can be analyzed using Mayer expansion \cite{Mayer-one,Mayer-two, PolyakovMayer}. The subtlety with the ${\ve}_2 \to 0$ limit is the clustering of the instanton particles, which leads to the multiple vertices of a given valency, labelled by an arbitrary positive integer $k$ (the number of instantons in a given cluster) weighted
by the partition function of a simpler one-dimensional gas
\[
\frac{1}{k! {\ve}_{2}^k} \int_{{\bR}^{k}} {\de} \left( 
\sum_{I} {\phi}_{I} \right)
{\dd}^{k} {\phi}\, 
\prod_{1 \leq I < J \leq k}
\frac{{\phi}_{IJ}}{{\phi}_{IJ} + {\ve}_{2}}
\]
which is equal to $\frac{1}{k^2 {\ve}_{2}}$, cf. \cite{MNSbound}. By summing over $k$'s
one arrives at the dilogarithm function in (\ref{eq:tbaii}). Finally, the limit
in (\ref{eq:wform}) singles out the tree diagrams, which lead to (\ref{eq:tbaii}), (\ref{eq:tbaiii}) with the propagator function $G(x)$ given by:
\[ G(x) = {\rm Limit}_{{\ve}_2 \to 0} \, \frac{{\CDd}(x) -1 }{{\ve}_2}
\]

\subsection{The sum over partitions}

The $U(N)$ instanton partition function can also be written as a sum over $N$-tuples
of partitions ${\la}^{(1)}, \ldots , {\la}^{(N)}$, which represent
various configurations of instantons sitting on top of each other in ${\bR}^4$. This representation can be effectively used to compute the first few terms in the ${\qq}$-expansion of ${\CW}^{\rm inst}$, or the first terms in the $\frac{1}{a^2}$ expansion, say, in the region $| a_{i} - a_{j} | \gg | {\ve} |$:
\[
{\CW}^{\rm inst} ( a, {\ve} ; m , {\qq}) \, =\,  \frac{N m(m+{\ve})}{\ve} {\varphi}({\qq})\ + \qquad\qquad\qquad \]
\begin{equation}
\qquad\qquad\qquad + \frac{1}{\ve}
\sum_{i \neq j} \frac{m^2 (m+{\ve})^2}{\left( a_{i}-a_{j} \right)^2 - {\ve}^2} \, 
{\qq} \frac{d {\varphi}({\qq})}{d{\qq}}  + \ldots
\label{eq:cwexp}
\end{equation}
where
\begin{equation}
{\varphi}({\qq}) = {\rm log} \prod_{n=1}^\infty \frac{1}{1- {\qq}^n}
\label{eq:dedekind}
\end{equation}

\newpage
\section{Discussion}

To conclude
we briefly review the relation of our results
to some recent work.  

{\bf The periodic Toda system:}
In \cite{KL}
the wavefunctions of the periodic Toda chain are constructed
in the form of integrals involving the solution $Q(x)$
of the Baxter equation. In order for the wavefunction
to belong to the $L^2$ Hilbert space the zeroes ${\de}_{1},
\ldots , {\de}_{N}$ of $Q$
must obey the {\sl quantization condition} 
(in \cite{KL}
the solution of Baxter equation is divided by the periodic
function 
$\prod_{i=1}^{N} {\rm sinh} \left( \frac{x - {\de}_{i}}{\ve} \right)$).   
We claim that the {\sl quantization conditions}
of \cite{KL} are equivalent to our Bethe equations for the pure ${\CN}=2$ $U(N)$ gauge theory, with the
identification ${\de}_{i} = a_{i}$.
We have therefore provided the  YY function for the {\it quantization
conditons in the periodic Toda chain}.  
Note that our perturbative equation 
(\ref{eq:smattoda}) 
is derived for the periodic Toda system in \cite{Fateev} from the approximate analysis of the Baxter equation. The observation of \cite{Fateev} was very helpful in relating
our two dimensional story \cite{BAsh2008-one, BAsh2008-two} to the four dimensional one. 

In \cite{Braverman} a {\it quasimap} partial 
compactification of the moduli
space of holomorphic maps of a sphere into the affine flag variety
$LG/T$ is studied, and the corresponding $J$-function \cite{Givental}
is shown to obey tautologically a non-stationary version of the periodic
Toda equation, the affine analogue of the result \cite{GiventalToda} for the $G/T$ type A sigma model. The intersection homology methods of \cite{Braverman}
are not applicable, it seems, to other quantum integrable systems. The questions addressed in the current paper do not appear in \cite{Braverman}, while the interesting setup of \cite{Braverman} has its own place (\ref{eq:kpsp}) in our general story, having to do with the four dimensional gauge theory in the presence of defects 
\cite{NekrasovIAS}, which we discuss elsewhere \cite{Tobe}. 
 
{\bf The elliptic Calogero-Moser system:}
In \cite{FV} the quantum $N$-particle elliptic Calogero-Moser system for the
$A_{N-1}$ system, for the integer coupling parameter ${\nu}$ was studied using the critical level
limit of the free field
representation of the
conformal blocks of the WZW conformal field theory
on a torus. The wavefunctions in \cite{FV} are written
in terms of 
\[
m = \frac{N(N-1)}{2}\left( {\nu}-1\right)
\] parameters 
$t_{{\al}_{i}}^{(i)}$, ${\al}_{i} = 1, \ldots , i({\nu}-1)$,
$i = 1, \ldots , N-1$, 
obeying the {\it elliptic Bethe-like} equations, which
are derived from the Yang-Yang function 
\[
Y( t_{\al}^{(i)}; {\xi}_{1}, \ldots , {\xi}_{N}) = 
2{\pi}i 
\sum_{i=1}^{N} 
\left( \frac{1}{2}{\tau}{\xi}_{i}^{2} + 
{\xi}_{i} \sum_{{\al}=1}^{i ( {\nu}-1)}
t_{\al}^{(i)} \right) -  S(t_{\al}^{(i)}; {\tau}) 
\]
with the quasimomenta ${\xi}_{1}, \ldots , {\xi}_{N}$
being the fixed parameters. One can supplement
the Bethe equations of \cite{FV} by the 
{\sl quantization conditions} ${\xi}_{1} = {\vartheta} + 
n_{1}, {\xi}_{2} = {\vartheta} + n_{2}, \ldots , 
{\xi}_{N} = {\vartheta} + n_{N}$, with $n_{i} \in 
{\bZ}$.
 We conjecture 
 \begin{equation}
 {\rm Crit}_{t_{\al}^{(i)}, \sum_{{\al}=1}^{i ({\nu}-1)}
 t_{\al}^{(i)} = a_{i}} \, 
S ( t_{\al}^{(i)} ; {\tau} ) = {\tilde W}^{\rm eff}
(a_{1}, \ldots , a_{N}; - {\nu}{\ve}, {\ve}, {\tau})
\label{eq:fvweff}
\end{equation}
Analogously, the critical value at fixed $\xi$ gives,
conjecturally, the Yang-Yang function for the type
B model. Note that the potential $S(t ; {\xi}; {\tau})$ of \cite{FV} corresponds, 
in the table of dualities sketched in \cite{BAsh2008-one, BAsh2008-two}
to ( a limit of) the four dimensional quiver gauge theory with the gauge group
$U \left( {\nu}-1 \right) \times 
U  \left( 2 \left({\nu}-1\right) \right) \times \ldots \times 
U \left( \left(  N-1\right) \left( {\nu}-1 \right)\right)$, 
compactified on the elliptic curve $E_{\tau}$.  The equivalence (\ref{eq:fvweff}) suggests an interesting
duality between the ${\CN}=2^*$ theory in the special $\Omega$ background
with the coupling $\tau$ and the quiver gauge theory. 

In \cite{L} a partial resummation of the
quantum mechanical perturbation theory around
the trigonometric
Sutherland model is proposed, leading to an
equation on the eigenvalue of the ${\hat H}_{2}$
operator of the elliptic Calogero-Moser system, for an arbitrary
value of the coupling ${\nu}$. 
The relation of this approach to the 
YY-function formalism is not
clear to us at the moment.

{\bf Parallel developments:}  There are several recent developments
 which we feel are related to our story,
 including the obvious ones, such as the
 integrability in the AdS/CFT context,
 but also \cite{WittenKapustin, WittenGukov, KontsevichSoibelman, AldayMaldacena,  GaiottoMooreNeitzke, AGT}. It is tempting to conclude from the convergence of these topics that the planar ${\CN}=4$ super-Yang-Mills is equivalent to the union of the vacuum sectors of all gauge theories with eight supercharges (${\CN}=2$ in four dimensions). 

We would like to return 
to all these questions in the future.

\bibliographystyle{ws-procs975x65}
\bibliography{ws-pro-sample}

\begin{thebibliography}{9}





\bibitem{MNS} G.~Moore, N.~Nekrasov, S.~Shatashvili, {\it Integration over the Higgs branches,} 
Comm.  Math. Phys. 209 (2000) 97-121, arXiv:hep-th/9712241.

\bibitem{GS-one}   
A.~Gerasimov, S.~Shatashvili, {\it Higgs Bundles, Gauge Theories and Quantum 
Groups,} Comm. Math. Phys. 277 (2008) 323-367, arXiv:hep-th/0609024.

\bibitem{GS-two}   

A.~Gerasimov, S.~Shatashvili, {\it Two-dimensional Gauge Theories and Quantum Integrable Systems},  In ``From Hodge Theory to Integrability and TQFT:  tt*-geometry", pp. 239-262, R. Donagi and K. Wendland, Eds., Proc. of Symposia in 
Pure Mathematics Vol. 78, American Mathematical Society, Providence, Rhode Island, 2008; arXiv:0711.1472.

\bibitem{BAsh2008-one} 

N.~Nekrasov, S.~Shatashvili,  {\it  Supersymmetric vacua and Bethe ansatz,} In ``Cargese 2008, Theory and Particle Physics: the LHC perspective and beyond", arXiv:0901.4744.

\bibitem{BAsh2008-two}

N.~Nekrasov, S.~Shatashvili, {\it Quantum integrability and supersymmetric vacua,} Prog. Theor. Phys. Suppl. 177:105-119, 2009, arXiv:0901.4748.

\bibitem{Neksw} N.~Nekrasov, {\it Seiberg-Witten prepotential from instanton counting,}  Adv. Theor. Math. Phys. 7:831-864, 2004,  hep-th/0206161.  

\bibitem{LNStest} A.~Losev, N.~Nekrasov, S.~Shatashvili, {\it Testing Seiberg-Witten solution,} 
In ``Cargese 1997, Strings, branes and dualities'' 359-372,  hep-th/9801061.

\bibitem{MNSbound} G.~Moore, N.~Nekrasov, S.~Shatashvili, {\it D particle bound states and generalized instantons,} Commun.  Math. Phys. 209 (2000) 77-95,  hep-th/9803265.

\bibitem{SW} N.~Seiberg, E.~Witten, {\it  Monopole Condensation, And Confinement In ${\CN}=2$ Supersymmetric 
Yang-Mills Theory,}
arXiv:hep-th/9407087, Nucl.~Phys. B426 (1994) 19-52; 
Erratum-ibid.B430:485-486, 1994 

\bibitem{GorskySW} A.~Gorsky, I.~Krichever, A.~Marshakov, A. Mironov, A. Morozov,  {\it Integrability and Seiberg-Witten exact solution,}    Phys. Lett. B355:466-474,1995,  hep-th/9505035.

\bibitem{WittenDonagi} R.~Donagi, E.~Witten, {\it Supersymmetric Yang-Mills theory and integrable systems,} Nucl. Phys. B460:299-334, 1996, hep-th/9510101. 

 \bibitem{YangYang} C. N. Yang, C. P. Yang, {\it Thermodinamics of a one-dimensional system of bosons 
with repulsive delta-function interaction,} J.  Math.  Phys. 10 (1969), 1115. 

\bibitem{Tobe}
N.~Nekrasov, S.~Shatashvili, 
{\it Supersymmetric vacua and quantum integrability},  to 
appear.
 


 \bibitem{HoriBook} K.~Hori, S.~Katz, A.~Klemm, R.~Pandharipande, R.~Thomas, 
 C.~Vafa, R.~Vakil, E.~Zaslow  (Eds.), {\it Mirror symmetry}, American Mathematical Society, Providence, 2003, 929p.  

 \bibitem{LNS}
A.~Losev, N.~Nekrasov, S.~Shatashvili, {\it  Issues in topological gauge theory,} Commun. Math. Phys. 209:97-121, 2000, 
hep-th/9711108.    

 \bibitem{Bethe} H.~Bethe, {\it On the theory of metals. 1. Eigenvalues and eigenfunctions for the linear atomic chain}, Z. Phys. 71: 205-226, 1931. 

 \bibitem{LMN} A.~Losev, A.~Marshakov, N.~Nekrasov, {\it  Small instantons, little strings and free fermions,} In ``Shifman, M. (ed.) et al.: From fields to strings, vol. 1",  581-621, hep-th/0302191.

\bibitem{MN} A.~Marshakov, N.~Nekrasov, {\it  Extended Seiberg-Witten Theory and Integrable Hierarchy}, JHEP 0701:104, 2007, hep-th/0612019.


\bibitem{KL} S.~Kharchev, D.~Lebedev, {\it
 Integral representations for the eigenfunctions of quantum open and periodic Toda chains from QISM formalism},
arXiv:hep-th/0007040,  J.~Phys.~A34 (2001) 2247-2258 

\bibitem{FV}  G.~Felder, A.~Varchenko,
{\it Integral representation of solutions of the elliptic Knizhnik--Zamolodchikov--Bernard equations}, arXiv:hep-th/9502165, Int.~Math.~Res.~Notices (1995) 221-233.

\bibitem{KricheverBook} H.~Braden, I.~Krichever, eds. {\it The Seiberg-Witten and Whitham Equations}, Gordon and Breach Science Publishers, 2000.



\bibitem{NOsw} N.~Nekrasov, A.~Okounkov, {\it Seiberg-Witten theory and random partitions}, hep-th/0306238.

\bibitem{TBAzam} Al.~Zamolodchikov, {\it
Thermodynamic Bethe Ansatz in Relativistic
Models. Scaling 3-state Potts and Lee-Yang Models}, 
Nucl. Phys. B342 (1990) 695.

\bibitem{Destri} C. Destri, H. J. de Vega, {\it New approach to thermal Bethe Ansatz}, Phys. Rev. Lett. 69 (1992) 
2313.

\bibitem{TBAblz} V.~Bazhanov, S.~Lukyanov and 
A.~Zamolodchikov, {\it Integrable
quantum field theories in finite volume: excited state energies}, Nucl. Phys.  B489 (1997) 487-531, hep-th/9607099.

\bibitem{TBAdor} P. ~Dorey and R.~Tateo, {\it Excited states by analytic continuation of TBA equations}, Nucl. Phys. B482 (1996) 639-659, hep-th/9607167.

\bibitem{TBAteschner} J.~Teschner, {\it On the spectrum of the sinh-Gordon model in finite volume}, hep-th/0702214.

\bibitem{Givental} A.~Givental, {\it Equivariant
Gromov-Witten Invariants}, 
arXiv:alg-geom/9603021 

\bibitem{GiventalToda} A.~Givental, {\it  Stationary Phase Integrals, Quantum Toda Lattices, Flag Manifolds and the Mirror Conjecture }, arXiv:alg-geom/9612001

\bibitem{Faddeevetc} L.~Faddeev, E.~Sklyanin, L.~Takhtajan, ÒQuantum inverse problem methodÓ, Theor.
Math. Phys. 40:2 (1980) 688-706, Teor.~Mat.~Fiz. 40:194-220,1979 (in Russian)

\bibitem{FadTak} L.~Takhtajan, L.~Faddeev,  {\it The Quantum method of the inverse problem and the Heisenberg XYZ model,}
Published in Russ. Math. Surveys  34:11-68,1979, Usp. Mat. Nauk 34:13-63, 1979. 

\bibitem{FaddeevLH} L.~Faddeev, {\it 
How algebraic Bethe ansatz works for integrable model},
hepth/9605187

 
\bibitem{Zhenya} E.~Sklyanin,  {\it Separation of variables - new trends},
Prog. ~Theor.~Phys.~Suppl.  118 (1995) 35-60, e-Print: solv-int/9504001

 \bibitem{Baxter} R.~Baxter, {\it Exactly solved models in statistical mechanics}, 
 Academic Press, London 1982
 
 \bibitem{fedya} F. A. Smirnov,  {\it Structure of Matrix Elements in Quantum Toda Chain, J. Phys. A.  31
(1998) 8953,}  arXiv:math-ph/9805011



\bibitem{BeilinsonDrinfeld} A.~Beilinson, V.~Drinfeld, {\it
Quantization of Hitchin's integrable system and Hecke eigensheaves},
preprint (ca. 1995), http://www.math.uchicago.edu~arinkin/langlands/ 








\bibitem{KricheverCM} I.~Krichever, {\it
Elliptic solutions of the KadomtsevÐPetviashvili equation and integrable systems of particles}, Funkts. Anal. Prilozh., 14:4 (1980), 45Ð54  
\bibitem{Inosemtsev} V.~Inozemtsev, 
{\it The Finite Toda Lattices}, Comm.~Math.~Phys.121 (1989)
629-638

\bibitem{Lame} G.~Lam\'e,  {\it Sur les surfaces isothermes dans les corps homognes en Žquilibre de tempŽrature},  J. Math. Pures Appl. 2 (1837), 147Ð188.


\bibitem{NekHol} N.~Nekrasov, {\it Holomorphic bundles and many-body systems,} Commun. Math. Phys. 180 (1996) 587-604, hep-th/9503157.

\bibitem{NekGor} A.~Gorsky, N.~Nekrasov, {\it Elliptic Calogero-Moser System from Two Dimensional Current Algebra}, hep-th/9401021.
 
\bibitem{Gaiotto} D.~Gaiotto, 
{\it ${\CN}=2$ dualities},  arXiv:0904.2715v1
        

\bibitem{WittenKapustin} A.~Kapustin, E.~Witten, 
{\it
  Electric-Magnetic Duality And 
  The Geometric Langlands Program},
  arXiv:hep-th/0604151

\bibitem{RS-i}  S.~Ruijsenaars, H.~Schneider, 
{\it A new class of integrable systems and 
its relation to solitons}, Ann. Phys. (NY) 170 (1986) 370--405. 

\bibitem{RS-ii} S.~Ruijsenaars, {\it Complete integrability of relativistic Calogero-Moser systems and elliptic function identities},
Comm.~Math.~Phys.  110 (1987) 191--213. 


\bibitem{Macdo} I.~Macdonald, {\it Symmetric Functions and Hall 
Polynomials}, Oxford Science Publications, 1998

\bibitem{NekFD} N.~Nekrasov, {\it Five dimensional gauge theories and
the relativistic gauge theories}, arXiv:hep-th/9609219, Nucl.Phys. {\bf B}531 (1998) 323-344 
 
 \bibitem{ABCD} N.~Nekrasov, S.~Shadchin, 
 {\it ABCD of instantons}, Comm.~Math.~Phys. 252 (2004) 359-391 
, arXiv:hep-th/0404225


    
    
 
 
\bibitem{NekAmsterdam} N.~Nekrasov, {\it Random partitions, topological strings and Mayer expansion,} lectures at the University of Amsterdam conference ``On random Partitions and Topological Strings",  June 2005.

\bibitem{NekrasovIAS} N.~Nekrasov, {\it Gauge theory special functions:
$Z$-functions for instantons with defects}, lecture at the DARPA meeting, {\it 
Langlands program and gauge theory}, IAS, Princeton, March 2004


\bibitem{Mayer-one} J.~Mayer, M.~G.~Mayer,
{\it Statistical Mechanics}, New York 1940

\bibitem{Mayer-two} J.~Mayer, E.~Montroll, 
{\it Molecular distributions}, 
 J. Chem. Phys. {\bf 9} (1941) 2-16
 
 \bibitem{PolyakovMayer} A.~Polyakov, 
 {\it Microscopic description of critical phenomena}
 JETP 55 (1968) 1026-1038



\bibitem{L} E.~Langmann, {\it  An explicit solution of the (quantum) 
    elliptic Calogero-Sutherland model}, 
    arXiv:math-ph/0407050


\bibitem{WittenGukov} S.~Gukov, E.~Witten, 
{\it Branes and Quantization}, 
 arXiv:0809.0305


\bibitem{Fateev}  C.~Ahn, V.~Fateev, C.~Kim, C.~Rim, B.~Yang, {\it Reflection Amplitudes of ADE Toda Theories and Thermodynamic Bethe Ansatz}, arXiv:hep-th/9907072, Nucl.Phys. B565 (2000) 611-628 


\bibitem{Braverman} A.~Braverman, {\it Instanton counting using affine Lie algebras I: equivariant $J$-functions of (affine) flag varieties and Whittaker vectors}, 
arXiv:math/0401409
 
 \bibitem{KontsevichSoibelman} M.~Kontsevich, Y.~Soibelman, {\it Stability structures, motivic Donaldson-Thomas invariants and cluster transformations}, arXiv:0811.2435

 
 \bibitem{AldayMaldacena}
L.~Alday, J.~Maldacena, {\it Null polygonal Wilson loops and minimal surfaces in Anti-de-Sitter space}, arXiv:0904.0663 
    


\bibitem{GaiottoMooreNeitzke}
D.~Gaiotto, G.~Moore, A.~Neitzke, 
{\it Wall-crossing, Hitchin Systems, and the WKB Approximation}, arXiv:0907.3987
 
 \bibitem{AGT} L.~Alday, D.~Gaiotto, 
 Y.~Tachikawa, {\it Liouville Correlation Functions from Four-dimensional Gauge Theories},  arXiv:0906.3219 
 
 \end{thebibliography}

\end{document}